\documentclass{article}
\RequirePackage[OT1]{fontenc} %IMS
\usepackage{parskip}

\usepackage{float}
\usepackage{titlesec}
\usepackage[slantedGreek]{mathpazo}
\usepackage[pdftex]{graphicx}
\usepackage{amsfonts}
\usepackage{amssymb}
\usepackage{amsthm}
\usepackage{graphicx}
\usepackage{amsmath}
\usepackage{mathtools}
\usepackage{lscape}
\usepackage{subcaption}
\usepackage{suffix}
\usepackage{nicefrac}

\usepackage{tikz}
\usetikzlibrary{matrix} % note this is in the preamble

\usepackage{booktabs}
\usepackage{longtable}
\usepackage{array}
\usepackage{multirow}
\usepackage{wrapfig}
\usepackage{float}
\usepackage{colortbl}
\usepackage{pdflscape}
\usepackage{threeparttable}
\usepackage{threeparttablex}
\usepackage[normalem]{ulem}
\usepackage{makecell}
\usepackage{microtype}

\usepackage{color}
\definecolor{DarkBlue}{rgb}{0.1,0.1,0.5}
\definecolor{Red}{rgb}{0.9,0.0,0.1}
\definecolor{Navy}{rgb}{0.00,0.00,0.30}
\definecolor{Yellow}{rgb}{1.00,1.00,0.00}
\definecolor{Gold}{rgb}{1.00,0.84,0.00}
\definecolor{Lightgoldenrod}{rgb}{0.93,0.87,0.51}
\definecolor{Goldenrod}{rgb}{0.85,0.65,0.13}
\definecolor{Black2}{rgb}{0.00,0.00,0.00}
\definecolor{orange}{rgb}{0.85,0.65,0.13}
\definecolor{SkyBlue}{rgb}{0.941176,0.972549,1.}
\definecolor{MyLightMagenta}{cmyk}{0.1,0.8,0,0.1}
\usepackage{xcolor}
\makeatletter
\def\mathcolor#1#{\@mathcolor{#1}}
\def\@mathcolor#1#2#3{%
  \protect\leavevmode
  \begingroup
    \color#1{#2}#3%
  \endgroup
}
\makeatother

\RequirePackage[colorlinks,citecolor=blue,urlcolor=blue]{hyperref} %IMS
\usepackage[square,sort,comma]{natbib}

\allowdisplaybreaks

\newcommand{\p}{{\bf p}}
\newcommand{\btheta}{\boldsymbol{\theta}}

\newcommand{\ben}{\begin{enumerate}}
\newcommand{\een}{\end{enumerate}}
\newcommand{\beq}{\begin{equation}}
\newcommand{\eeq}{\end{equation}}
\newcommand{\bde}{\begin{description}}
\newcommand{\ede}{\end{description}}

\newcommand{\abs}[1]{\lvert#1\rvert}

\newcommand{\NormRV}{\mathcal{N}}

\newcommand{\UnifRV}{\mathcal{U}}

\newcommand{\Cov}{\mathrm{Cov}}
\newtheoremstyle{slplain}% name
  {1\baselineskip\@plus.2\baselineskip\@minus.2\baselineskip}% Space above
  {.5\baselineskip\@plus.2\baselineskip\@minus.2\baselineskip}% Space below
  {\slshape}% Body font
  {}%Indent amount (empty = no indent, \parindent = para indent)
  {\bfseries}%  Thm head font
  {.}%       Punctuation after thm head
  { }%      Space after thm head: " " = normal interword space;
  {}%       Thm head spec

\theoremstyle{slplain}

\newtheorem{theorem}{Theorem}

\newtheorem{definition}[theorem]{Definition}
\newtheorem{example}[theorem]{Example}

\usepackage{booktabs,array}

\newcount\rowc

\newcommand{\ind}{\stackrel{\mathrm{ind}}{\sim}}

\graphicspath{{../../art/}{../art/}{./art/}{./}}
\graphicspath{{../../figures/}{../figures/}{./figures/}{./}}

\newcommand{\E}{\mathbb E}
\newcommand{\V}{\mathbb V}

\numberwithin{equation}{section}

\usepackage[inline]{enumitem}
\setlist*[enumerate]{label=(\roman*)}

\title{Inverse Probability Weighting: from Survey Sampling to Evidence Estimation \thanks{This is a preprint of a manuscript currently under peer review.}}
\author{Jyotishka Datta \\ Department of Statistics \\ Virginia Tech \and Nicholas G. Polson \\ Booth School of Business \\ University of Chicago}
\date{}

\begin{document}
\maketitle
% \begin{frontmatter}
% \title{Inverse Probability Weighting: the Missing Link between Survey Sampling and Evidence Estimation\thanksref{T1}}
% \runtitle{Inverse Probability Weighting: the Missing Link}

% \begin{aug}
% \author{\fnms{Jyotishka} \snm{Datta}\thanksref{addr1}\ead[label=e1]{jyotishka@vt.edu}}
% \and
% \author{\fnms{Nicholas} \snm{Polson}\thanksref{addr2}\ead[label=e2]{ngp@chicagobooth.edu}},

% \runauthor{Datta and Polson}

% \address[addr1]{Department of Statistics, Virginia Polytechnic Institute and State University, Blacksburg, VA 24061. E-mail: \printead{e1}.
% }
% \address[addr2]{ Booth School of Business, University of Chicago. E-mail: \printead{e2}.
% }

% \end{aug}

\begin{abstract}
We consider the class of inverse probability weight (IPW) estimators, including the popular Horvitz--Thompson and H\'ajek estimators used routinely in survey sampling, causal inference and evidence estimation for Bayesian computation. We focus on the `weak paradoxes' for these estimators due to two counterexamples by \citet{basu1988statistical} and \citet{wasserman2004bayesian} and investigate the two natural Bayesian answers to this problem: one based on binning and smoothing : a `Bayesian sieve' and the other based on a conjugate hierarchical model that allows borrowing information via exchangeability. We compare the  mean squared errors for the two Bayesian estimators with the IPW estimators for Wasserman's example via simulation studies on a broad range of parameter configurations. We also prove posterior consistency for the Bayes estimators under missing-completely-at-random assumption and show that it requires fewer assumptions on the inclusion probabilities. We also revisit the connection between the different problems where improved or adaptive IPW estimators will be useful, including survey sampling, evidence estimation strategies such as Conditional Monte Carlo, Riemannian sum, Trapezoidal rules and vertical likelihood, as well as average treatment effect estimation in causal inference.
\end{abstract}


% \begin{keyword}
% \kwd{Inverse probability weighting}
% \kwd{Horvitz--Thompson}
% \kwd{H\'ajek}
% \kwd{Importance sampling}
% \kwd{Stein phenomenon}
% \kwd{Bias-variance trade-off}
% \end{keyword}

% \end{frontmatter}

% {\it Keywords:}  
% Instrumental Variable ; Causal Inference. 

\section{Introduction}

Inverse probability weight (IPW) estimators have been used across statistical literature in diverse forms: in survey sampling, in designing importance sampling in Monte Carlo techniques and in the context of average treatment effect estimation in causal inference. In survey sampling, the goal is often to estimate population mean $\psi$ from a finite sample ($y_1, \ldots, y_n)$, and a common approach is to weigh each observation $y_i$ by a weight $w_i$ inversely related to their probability of inclusion (probability proportional to selection, or PPS). For example, common PPS estimators admit the form $\hat{\psi} = \sum_{i \in s} w_i y_i / n$ or a `ratio' estimator: $\hat{\psi} = \sum_{i \in s} w_i y_i / \sum_{i \in s} w_i$, where $s$ denotes the sample, and $w_i$'s denote the sampling weights. These weights or probabilities of inclusion might be known, or unknown depending on whether their source is sampling design or non-response. Two of the most popular examples of such estimators are the Horvitz--Thompson estimator \citep{horvitz1952generalization}, which uses fixed weights to provide an unbiased estimate of the population mean, and the H\'ajek estimator \citep{hajek1971comment}, which normalizes the weights to improve stability at the cost of introducing slight bias.

Similarly, in importance sampling, the goal is to estimate the evidence $\psi = \E_{F}\{l(X)\} = \int l(x) dF(x)$ where $F$ might be difficult to sample from and a common technique is to estimate $\psi$ by drawing samples from a candidate distribution $G$ (with density $g$) and calculate:
\[
\hat{\psi} = n^{-1} \sum_{i=1}^{n} l(x_i) f(x_i)/g(x_i) \doteq n^{-1} \sum_{i=1}^{n} l(x_i) w(x_i).
\]
The candidate density $g(\cdot)$ is chosen such that the `weights' $w(x_i)$ is nearly constant \citep{firth2011improved}. The `ratio' estimator analog in importance sampling would correspond to $\sum_{i=1}^{n} l(x_i) w(x_i)/\sum_{i=1}^{n}w(x_i)$ as developed by \citet{hesterberg1988advances, hesterberg1995weighted}, and further developed in \citep{firth2011improved}. The unattainable `optimal' choice of $w(\cdot)$ is of course $f(\cdot)$ itself,  and a key insight in producing more accurate estimation is that self-normalization or biasing the sampler towards low probability regions can help. Methods such as nested sampling or vertical likelihood use a Lorenz curve re-ordering of summation to achieve this goal \citep{polson2014vertical, chopin2010properties, skilling2006nested, datta2023quantile}. 

% Another alternative, and largely overlooked, approach for Monte Carlo estimation based on Riemann sums or trapezoidal rule was proposed by \citet{yakowitz1978weighted} and generalized by \citet{philippe1997processing,philippe2001riemann} that drastically improves the convergence rate by reducing the variance. These methods achieve convergence rates of $O(n^{-4})$ for \citet{yakowitz1978weighted} and $O(n^{-2})$ for \citet{philippe1997processing} improving on the usual $O(n^{-1})$ rate of Monte Carlo estimation by standard averaging. 

In this paper, we first contrast and compare the popular inverse probability weight estimators—Horvitz--Thompson and H\'ajek—and discuss their relative merits and demerits in light of two weak paradoxes, viz., Basu's circus example \citep{basu1988statistical} and the Robins–Ritov–Wasserman example \citep{robins1997toward, wasserman2004bayesian}.  We then discuss a binning-and-smoothing estimator by \citet{ghosh2015weak} and a hierarchical Bayes estimator by \citet{li2010bayesian}, in the light of the \citet{wasserman2004bayesian}'s example, and show how they can lead to a possible resolution. We also build connections between IPW estimators and evidence estimation techniques and argue that these innovative ideas from Monte Carlo methods can be exploited in designing IPW estimators to achieve a lower variance and higher stability. We argue that the issue of choice of weights -- an optimal proposal $g(\cdot)$ in Monte Carlo or sampling weights -- provides new insights on a long-standing controversy about a `weakness' in Bayesian paradigm \citep[][e.g.]{wasserman2004bayesian, sims2010understanding, ghosh2015weak, li2010bayesian}. 

There is a large, influential literature on statistical methods for improving survey-weighted estimates, particularly, in the context of complex survey designs. Our goal is not to attempt a review of these methodological advances in survey weight regularization or calibration, and we point the interested reader to the comprehensive review by \citet{chen2017approaches} and the references therein, e.g. \citet{haziza2017construction}. \citet{haziza2017construction} provides a comprehensive review of methods for constructing survey weights, focusing on techniques to adjust for unequal probabilities of selection, nonresponse, and post-stratification to improve the representativeness and accuracy of survey estimates. \citet{chen2017approaches} discuss the limitations of basic design-based weights, derived from the inverse of inclusion probabilities, and propose modifications such as weight trimming, weight modeling, and incorporating weights into statistical models. 

The structure of this article is as follows. In \S \ref{sec:ipw}, we define the popular IPW estimators and recent developments, and their connections with importance sampling. In \S \ref{sec:paradox}, we discuss two popular examples of `weak paradoxes' due to Basu, and Robins-Ritov-Wasserman. For the latter, we derive asymptotic properties of a Bayes estimator due to \citet{li2010bayesian} in \S \ref{sec:theory}. We compare different survey sampling estimators through a range of simulation studies in \S \ref{sec:sim}, to illustrate their relative merits and demerits. Finally, in \S \ref{sec:dis}, we discuss other areas of connection such as average treatment effect estimation in potential outcomes framework, and suggest a few possible future directions. 

\section{Inverse Probability Weight Estimators}\label{sec:ipw}

\subsubsection*{Horvitz--Thompson and Hájek Estimators}

We begin by defining two widely used estimators in survey sampling: the Horvitz--Thompson (HT) estimator \citep{horvitz1952generalization} \footnote{Also called Narain-Horvitz-Thompson estimator after \citet{narain1951sampling} by \citet{rao1999some, chauvet2014note}.} and the Hájek estimator \citep{hajek1971comment} briefly. Consider a finite population $U = \{1, 2, \dots, N\}$ with values $Y_k$ for each unit $k \in U$. A sample $s \subset U$ is drawn according to a sampling design with known inclusion probabilities $p_k = P(k \in s)$. \footnote{In a probability proportional to size (PPS) design, where the inclusion probabilities are proportional to an auxiliary variable $x_k$, they take the form
\begin{equation}
    p_k = \frac{n x_k}{\sum_{i=1}^{N} x_i}.
\end{equation}}
The Horvitz--Thompson estimator for the population mean $\psi = N^{-1} \sum_{k \in U} Y_k$ is then
\begin{equation}
    \hat{\psi}_{HT} = \frac{1}{N}\sum_{k \in s} \frac{Y_k}{p_k}.\quad \text{(Horvitz--Thompson)} \label{eq:ht}
\end{equation}
The H\'ajek estimator is an alternative procedure that estimates the population sum as well as $N$ i.e., normalizes by the estimated total,
\begin{equation}
    \hat{\psi}_{\text{H\'ajek}} = \frac{\sum_{k \in s} Y_k /p_k}{\sum_{k \in s} 1/p_k}.\quad \text{(H\'ajek)} \label{eq:hajek} 
\end{equation}
The HT estimator \eqref{eq:ht} is an unbiased estimator of $\psi$, while the H\'ajek estimator \eqref{eq:hajek} is non-linear, and approximately unbiased estimator that does not require the knowledge of population size $N$, irrespective of whether $N$ is known.  

\subsubsection*{Missing Data Framework and Adaptive normalization} 

An equivalent framework arises in the context of missing data, \textit{i.e.}, when subjects are observed with nonuniform probabilities. We follow the notations of \citet{khan2023adaptive} to describe it here. Here our goal is to estimate the mean $\psi$ of $Y_1, \ldots, Y_n$, with the observations missing at random. The Bernoulli indicators $R_k, k = 1, \ldots, n$ indicate whether $Y_k, k = 1, \ldots, n$ were observed or not. We assume $R_k \ind \text{Bernoulli}(p_k)$ for $k = 1, \ldots, n$, which reflects non-response bias in sample surveys.  Then, the Horvitz--Thompson and H\'ajek estimator of $\psi = n^{-1} \sum_{k=1}^n Y_k$ are:
\begin{align}
    \hat{S} = \sum_{k=1}^{n}\frac{Y_k R_k}{p_k} &, \quad \hat{n} = \sum_{k=1}^{n}\frac{R_k}{p_k} \label{eq:ipw} \\
    \hat{\psi}_{HT} = \frac{\hat{S}}{n}, &\quad     \hat{\psi}_{\text{H\'ajek}} = \frac{\hat{S}}{\hat{n}}. \label{eq:ipw-2}
\end{align}

A notable generalization of the aforementioned estimators is the Trotter--Tukey estimator \citep{trotter1956conditional}. Recently, \citet{khan2023adaptive} rediscovered the Trotter--Tukey estimator as the adaptive normalization (AN) idea by letting data choose the tuning parameter $\lambda$. 
\begin{align}
    \hat{\psi}_{TT} & = \frac{\hat{S}}{(1-\lambda)n + \lambda \hat{n}}, \; \lambda \in \mathbb{R} \quad \text{(Trotter--Tukey)} \label{eq:tt} \\ 
    \hat{\psi}_{AN} & = \frac{\hat{S}}{(1-\hat{\lambda})n + \hat{\lambda} \hat{n}}. \; \quad \text{(Adaptive Normalization)} \label{eq:an}
\end{align}
Furthermore, \citet{khan2023adaptive} show that the `adaptive normalization' idea dating back to \citep{trotter1956conditional} leads to a lower asymptotic variance than both while generalizing these estimators, in particular, $\V(\hat{\psi}_{AN})$ is lower than both $\V(\hat{\psi}_{HT})$ as $\V(\hat{\psi}_{\text{H\'ajek}})$. \citet{khan2023adaptive} also provide empirical evidence that their adaptive normalization scheme leads to a lower mean squared error of IPW estimators in different application areas in average treatment effect estimation and policy learning. 

\subsection{Horvitz-Thompson vs. H\'ajek Estimator}\label{sec:ht}
Horvitz--Thompson estimators possess many desirable properties: they are unbiased, admissible and consistent. \citet{ramakrishnan1973alternative} provides a simple proof that the HT estimator is admissible in a class of all unbiased estimators of a finite population total. \citet{isaki1982survey} proves that they achieve consistency under suitable conditions, such as, specifically requiring boundedness of population and weight sequences, bounded product of inclusion probabilities and second moments, and a variance that is \( O(n_t^{-1}) \). \citet{delevoye2020consistency}  provides conditions such as boundedness of moments for the outcome and inclusion probabilities and weak-design dependence. \citet{delevoye2020consistency} also point out that these conditions might be violated in ill-behaved setting with heavy-tailed outcomes or skewed sampling designs, common in natural experiments, where the practitioner has little control over the design. 

It is worth noting here that IPW estimators, in particular the HT estimator, can be looked at as a weighted estimator, where the weights are not model-based, but design-based, as pointed out by several authors from \citet{neyman1934two} to \citet{little2008weighting}. For example, in the HT estimator \eqref{eq:ht}, the $k^{th}$ unit is assigned a weight proportional to the inverse of the selection probability as it `represents the $1/p_k$ units of the population. %We return to this point in \S \ref{sec:mc} to point out a semi-parametric model due to \citet{kong2003theory} yielding HT estimator as a special case of a fully exponential model. 

\subsubsection{When is the H'ajek estimator preferable?}
We follow \citet[ch.~5]{sarndal2003model} for this discussion. First, note that the two estimators, HT and H\'ajek, can produce identical results under certain designs. However, if the population total $N$ is unknown, only the H\'ajek estimator~\eqref{eq:hajek} can be used. On the other hand, if $N$ is known---as is typically the case in the finite population estimation problem---the H\'ajek estimator ``\textit{is usually the better estimator, despite estimation of an} \textit{a priori} \textit{known quantity}''~\citep{sarndal2003model}. \citet{sarndal2003model} supports this by identifying three situations where the weighted mean estimator (H\'ajek) outperforms the simple unbiased estimator (HT). In particular, \citet[p.~183]{sarndal2003model} shows that the $\hat{\psi}_{\text{H\'ajek}}$ estimator achieves lower variance than $\hat{\psi}_{\text{HT}}$ in each of these cases:

\begin{enumerate}
    \item When the values of $y_k$ are closer to the population mean $\psi$, a special case being fixed $y_k = c$ for $k = 1, \ldots, N$. 
    \item When the sample size $n_s$ is variable with equal inclusion probabilities, e.g. the case where $y_k = c$ and $p_k = p_0$ or $k = 1, \ldots, N$, but the sample sizes vary. 
    \item Finally, when the inclusion probabilities $p_k$ are negatively correlated with the $y_k$ values, with large $y_k$ values corresponding to small $p_k$ values and vice versa. It is easy to see that the H\'ajek estimator is adaptable to high fluctuations in $y_k/p_k$ values that HT will suffer from, leading to high variability. \citet{basu1988statistical} pointed this out in his famous circus example, recounted in \S \ref{sec:ht}. 
\end{enumerate}

Despite being popular choices, both the Horvitz--Thompson (HT) \citep{horvitz1952generalization}, and the H\'ajek estimator \citep{hajek1971comment} have generated debates and controversies in the recent past. For example, the Horvitz--Thompson estimator where we encounter randomly missing observations and a very high-dimensional parameter space is presented as evidence of weakness of Bayesian method in this problem, in an example due to \citet{wasserman2004bayesian}, based on \citet{robins1997toward}. In response, a few authors \citep{sims2010understanding, ghosh2015weak, li2010bayesian, harmeling2007bayesian} have tried to provide a Bayesian answer by constructing Bayesian estimators that achieve a lower variance than the HT estimator at the cost of admitting a small (and in some cases, vanishing) bias: another example of the bias-variance trade-off. 

\noindent \textbf{Stein's Paradox:} Here, we argue that this phenomenon is another example of Stein's paradox \citep{james1961estimation, efron1977stein, stigler19901988} observed in a different context. The original Stein's paradox showed the inadmissibility of the ordinary maximum likelihood estimator which is both unbiased and minimax for normal means in dimensions more than $3$ by shrinking each component towards the origin (or a pre-fixed value), thereby borrowing strength from each other. The Stein's paradox and the James--Stein shrinkage estimator has been truly transformative in statistics in both establishing Empirical Bayes' success in high-dimensional inference and inspiring both frequentist shrinkage estimators and Bayesian shrinkage priors in such problems. We argue that the simple HT estimator can be improved upon by borrowing strength by moving from an independence framework to an exchangeable one, exhibiting Stein's shrinkage phenomenon and effects of regularization. 

Before we discuss the weak paradoxes, we briefly discuss how the connection between IPW estimators and Monte Carlo sampling, in particular importance sampling, and various improvements such as Riemann sums \citep{philippe1997processing, philippe2001riemann} or vertical likelihood integration that applies `binning and smoothing' using a score-function heurism to choose the weight function \citep{polson2014vertical, madrid2018deconvolution}. We then show how the idea of binning and smoothing also improves the HT estimator \citep{ghosh2015weak} in the apparent weakness example due to \citet{wasserman2004bayesian}. %Finally, we discuss the implications of Stein's paradox and inadequacy of HT estimators in Causal inference framework, viz. in average treatment effect estimation \citep{rubin1974estimating, imbens2004nonparametric}. 
We discuss these connections briefly in the next subsection and point out the similarities between estimators employed in survey sampling and Monte Carlo integration and their connections with statistical mechanics. 

\subsection{Importance sampling}\label{sec:mc}
One way to look at this connection between sampling strategies and integral approximation is to represent the former as a missing data problem. Suppose that our goal is to estimate:

$$\psi = \int_0^1 y(x) dx,$$

which can be thought of as a limiting value of $\psi_n = n^{-1}\sum_{i=1}^{n}y_i$ as $n \to \infty$, and can be approximated up to any degree of precision by $\psi_N = N^{-1}\sum_{i=1}^{N}y_i$ for a sufficiently large $N$. Given a random sample of size $n \ll N$, with sample probabilities $\pi$ attached to $y_i$, we can view estimating $\theta$ as a problem of estimating the population quantity $\psi_N$ by $\psi_n$. The usual importance sampling estimator in this case is akin to using $\E_{\pi}[\sum_{i=1}^{n} y_i/\pi_i] = \psi$, the usual HT estimator. Just like the HT estimator, the variance of importance sampling could blow up for poor choices of $g(\cdot)$, while the bias may shrink to zero. As we show below, these connections have been exploited by several authors to propose alternative importance sampling strategies. 

Given a sample $(y_1, \ldots, y_n)$, from either the density $f$ corresponding to $F$ itself or a suitable proposal density $g$, the usual importance sampling estimates $\psi$ by the empirical average: 
\begin{equation}
    \hat{\psi}_{IS} = \frac{1}{n} \sum_{i=1}^{n} \frac{l(y_i) f(y_i)}{g(y_i)} \label{eq:is}.
\end{equation}
If the underlying probability measure is easy to sample from, \textit{i.e.} if we can afford $f = g$, the above will reduce to the empirical average $\hat{\psi} = n^{-1} \sum_{i=1}^{n} l(y_i)$, which by the Law of Large Numbers, converge to the true value of $\psi$ at $O(n^{-1})$ rate. It is worth pointing out that replacing the empirical average for the na\"ive importance sampling estimator in \eqref{eq:is} by a Riemann sum estimator provides a remarkable improvement in stability and convergence as demonstrated in \citep{philippe1997processing, philippe2001riemann} or \citep{yakowitz1978weighted}. Recently, \citet{datta2023quantile} showed that combining Riemann sum estimators with nested sampling ideas can yield sharper convergence rates (up to $O(n^{-4})$) for marginal likelihood estimation in Bayesian problems. We refer the reader to \citet{datta2023quantile} for a more detailed discussion of vertical likelihood, and how it connects importance sampling with nested inference. This framework also offers a unifying perspective on several sampling-based approaches that aim to improve the stability of likelihood-weighted estimators.

\subsubsection{A Bayesian perspective} We conclude the discussion of evidence estimation with a pertinent and profound point raised by \citet{diaconis1988bayesian} in support of Bayesian approaches for probabilistic numerical integration. \citet{diaconis1988bayesian} starts with a algebraic function:

\[
f(x) = \exp\left\{\cosh\left(\frac{x+2x^2+\cos x}{3 + \sin x^3} \right)\right\}, \quad f: [0,1] \mapsto \mathbb{R},
\]
for which we want $\int_0^1 f(x) dx$ and asks ``\emph{What does it mean to `know' a function?.}" In such a situation, where we might know a few properties of $f$ and not others, a Bayesian approach seems natural, where one starts with a prior on $\mathcal{C}[0,1]$, the space of continuous functions, and estimate the integral using Bayes' rule. \citet{diaconis1988bayesian} shows that Brownian motion, the `easiest' prior on  $\mathcal{C}[0,1]$, yields a linear interpolation leading to the trapezoid rule. \citet{diaconis1988bayesian} then shows that a host of well-known methods can be recovered as Bayesian estimates. This key question is revisited in \citep{owen2019comment} where he asks what does it mean to know the error $\Delta = \abs{\hat{\psi} - \psi}$ and discusses advantages of Bayesian approaches over classical methods. \citet{owen2019comment} argues in favour of Bayes methods in difficult problems where extreme cost of function evaluation or skewness or heavy-tailed properties of $l(x)$ or unavailability of central limit theorem exposes the weakness of classical methods. 

\section{Weak Paradoxes in Ratio Estimation Problem}\label{sec:paradox}

\subsection{\citet{basu1988statistical}'s example}

\begin{example}{\textbf{Basu's circus example:}} The famous circus example, due to \citep{basu1988statistical}, shows that the HT estimator \eqref{eq:ht} could lead to absurd estimates in some situations. Here, we imagine a circus owner, trying to estimate the total weight of his $50$ elephants (say $Y$), picks a representative elephant from his herd (\textit{Sambo}) and multiply his weight by $50$. But, then a circus statistician, appalled by this estimate, devices a plan where \textit{Sambo} is picked with probability $99/100$ and each of the remaining with $1/100$. Unfortunately, now the HT estimate is $\text{Sambo's weight}\times100/99$, a serious underestimate, and moreover, if the owner picks \textit{Jumbo}, the biggest in the herd, the HT estimate is $\text{Jumbo's weight}\times4900$, an absurdity.
\end{example} 

Discussing \citet{basu1988statistical}, \citet{hajek1971comment} points out that the HT estimator's {\em ``usefulness is increased in connection with ratio estimation"} and proposed an estimator in presence of auxiliary information $A_k$, related to $Y_k$ and with known total:
\begin{equation}
\hat{Y}_{\text{H\'ajek}} = \sum_{k=1}^{n}A_k \times \frac{\sum_{k=1}^{n}\frac{Y_k}{p_k}}{\sum_{k=1}^{n}\frac{A_k}{p_k}}, \; \text{where } Y_k \propto A_k, k = 1, \ldots, n, \label{eq:hajek-2} \\
\end{equation}
which would not be affected in this example like the standard HT estimator. The H\'ajek IPW estimate for the population mean in \eqref{eq:hajek} can be derived from the special case $A_k \equiv 1, \forall k$.  

Although the circus example was intended to be a pathological example, it provides at least two useful insights. First, weighted estimators can lead to nonsensical answers despite having nice large sample properties \citep{little2008weighting}. Second, problems of similar nature occur in importance sampling or Monte Carlo estimation of the marginal likelihood where empirical averages or unbiased estimator could have high or infinite variance \citep{li2010bayesian, raftery2006estimating}. As pointed out in \citep{datta2023quantile}, strategies like Riemann sum \citep{philippe1997processing, philippe2001riemann}, or careful choice of weight functions like vertical likelihood \citep{polson2014vertical} or nested sampling \citep{skilling2006nested}, or using ratio estimators \citep{firth2011improved}, or adaptive normalization \citep{ khan2023adaptive} can resolve these issues. In particular, the same trick of ordering the draws and applying a Riemann-sum type approach is the key to avoid falling into examples like Basu's circus. 

\subsection{\citet{wasserman2004bayesian}'s example:} We first re-state the example which is itself a simplification of an example from \citep{robins1997toward}, in a similar spirit. We consider IID samples $(Y_i, X_i, R_i)$, $i = 1, \ldots, B$ such that $Y_i$'s are generated as a mixture of of Bernoulli distributions with individual parameters indexed by the component label $X_i$, and the `missingness' indicator $R_i$ denoting whether $Y_i$ was observed or not. Let the `success' probabilities associated to $R_i$ be known constants $p_{X_i}$ satisfying:
\begin{equation}
    0 < \delta \le p_j \le 1 - \delta < 1, \; j = 1, \ldots, B. \label{eq:p-bound}
\end{equation}
Note that this strong condition on $p_j$'s in \eqref{eq:p-bound} ensures that the HT estimator will not lead to absurd answers like Basu's paradox stated earlier. The hierarchical model for each draw  $(Y_i, X_i, R_i)$, $i = 1, \ldots, n$, is:
\begin{align}
    X_i &\sim \UnifRV(1, \ldots, B) \\
    [R_i & \mid X_i = x_i] \sim \text{Bernoulli}(p_{x_i}) \\
    [Y_i &\mid R_i, X_i = x_i] \sim \begin{cases} \text{unobserved} \; \text{if} \; R_i = 0 \\
    \text{Bernoulli}(\theta_{x_i}) \; \text{if} \; R_i = 1.
    \end{cases} \label{eq:was}
\end{align}
The parameter of interest is the average $\psi = (1/B)\sum_{b=1}^B \theta_b$. \citet{wasserman2004bayesian} argues that since the likelihood has little information on most $\theta_j$'s and the known constants $B$ and $p_j$'s drop from the likelihood, Bayes' estimates for $\psi$ are going to be poor. On the other hand, the HT estimate:
\begin{equation}
    \hat{\psi}_{HT} = \frac{1}{n}\sum_{i=1}^n \frac{R_i Y_i}{p_{X_i}} \label{eq:psi_ht}
\end{equation}
is easily seen to be unbiased, given $\E(R_i Y_i/p_{X_i}) = \E\{ \E(R_i \mid Y_i, X_i) \cdot \E(Y_i \mid X_i) /p_{X_i}\} = \E(\E(Y_i \mid X_i)) = \psi$ since $\E(R_i \mid Y_i, X_i) = p_{X_i}$ by construction. The HT estimate will also satisfy $\V(\hat{\psi}_{HT}) \le 1/n\delta^2$, using Hoeffding's inequality. We would like to note here that the \textit{assumption} \eqref{eq:p-bound} that the selection probabilities are between $\delta$ and $1-\delta$ is exploited explicitly in this asymptotic result, \textit{i.e.,} the HT estimator might have infinite variance as $\delta$ goes to zero, making the assumptions crucial. 

This example of apparent weakness in Bayesian paradigm and the concluding remarks by \citet{wasserman2004bayesian} \footnote{``\emph{Bayesians are slaves to the likelihood function. When the likelihood goes awry, so will Bayesian inference.}" \citep[pp.189]{wasserman2004bayesian}} has since been a source of debate and elicited response from Bayesian community \citep{li2010bayesian,sims2010understanding,harmeling2007bayesian,ghosh2015weak} which we review briefly below. 

\subsection{Bayesian solutions for \citet{wasserman2004bayesian}'s problem}\label{sec:sol}

We review the existing Bayesian resolutions for the Robins-Ritov-Wasserman problem using Bayesian ideas and argue that, in some special cases, the improvement in accuracy of Bayes' estimate is attained via exploiting the `borrowing strength' phenomenon in Stein's shrinkage. 

\subsubsection{Full Bayes Solution \citep{li2010bayesian}}

\begin{figure}[ht!]
    \centering
\begin{tikzpicture}
\matrix[matrix of math nodes, column sep=20pt, row sep=20pt] (mat)
{
    & & \alpha_F & & &\\ 
    & & \alpha_T, \psi & &[4em] (p_1, \ldots, p_B) &\\ 
    (X_1, \ldots, X_B) & \theta_1 & \ldots & \theta_B & (R_1, \ldots, R_B) & \\
    % x_{1} & \ldots & x_{B} \\
    & y_{1} & \ldots & y_{B} &\\
    & & |[blue]| \psi & &\\
};

\foreach \column in {2, 4}
{
    \draw[->,>=latex] (mat-2-3) -- (mat-3-\column);
    \draw[->,>=latex] (mat-3-\column) -- (mat-4-\column);
    \draw[->,>=latex] (mat-3-1) -- (mat-4-\column);
    \draw[->,>=latex] (mat-3-5) -- (mat-4-\column);
    \draw[->,>=latex] (mat-4-\column) -- (mat-5-3);
}

\draw[->,>=latex] (mat-1-3) -- (mat-2-3);
\draw[->,>=latex] (mat-2-5) -- (mat-3-5);
% \draw[->,>=latex] (mat-3-5) -- (mat-4-4);

\node[anchor=east] at ([xshift =-40pt]mat-2-3) 
{$\mathcolor{red}{\psi \sim \text{Beta}(\alpha_F, \alpha_F)}$};
% \node[anchor=east] at ([xshift =-30pt]mat-2-3) 
% {$\theta_b \ind \text{Beta}(\alpha_T \psi, \alpha_T (1-\psi_0))$};
% \node[anchor=west] at ([xshift =40pt]mat-2-5){(\textcolor{red}{known} $\mathcolor{red}{p_i's}$)};
\end{tikzpicture}
    \caption{Hierarchical model for the Robins-Ritov-Wasserman problem}
    \label{fig:li}
\end{figure}
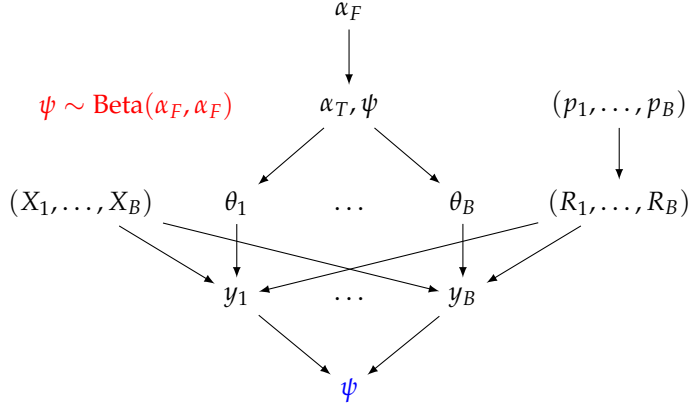

\citet{li2010bayesian} provides a simple Bayes estimator by assuming that $\theta_1, \ldots, \theta_B$ are exchangeable, not independent (see Fig. \ref{fig:li}). \citet{li2010bayesian} estimates $\psi$ by augmenting \eqref{eq:was} by with Beta priors for $\theta_1, \ldots, \theta_B$ and a further hyperprior on the mean of these Beta priors, as follows:
\begin{align}
    \theta_b \mid \alpha_T, \psi & \ind \text{Beta}(\alpha_T \psi, \alpha_T (1-\psi_0)) \\
    (\psi, \alpha_T \mid \alpha_F) & \sim \text{Beta}(\alpha_F, \alpha_F) \times \pi(\alpha_T),
\end{align}
where, $\psi$ is the mean of $\theta$, and the shape parameters $\alpha_T$ and $\alpha_F$ control the width of range of $\btheta$ and the concentration of $\psi$. 

Using this model, \citet{li2010bayesian} derives a posterior mean estimate of $\psi$ as:
\begin{equation}
  \hat{\psi}_{Li} = \frac{\sum_{i=1}^{n} R_i Y_i + \alpha_F}{\sum_{i=1}^{n} R_i + 2\alpha_F} \label{eq:li} %\asymp \frac{\sum_{i=1}^{n} R_i Y_i}{\sum_{i=1}^{n} R_i} 
\end{equation}

Through extensive numerical simulation, \citet{li2010bayesian} shows that this estimator achieves a smaller mean squared error compared to the HT estimator in \citet{wasserman2004bayesian}. \citet{li2010bayesian} also argue that the variance of HT estimator, given by:
\begin{equation}
    \V(\hat{\psi}_{HT}) = \frac{1}{n} \left\{ \frac{1}{B}\sum_{b=1}^{B}\frac{\theta_b}{p_b} - \psi^2 \right\}, \label{eq:v-ht}
\end{equation}
will lead to inflated variance for small $p_b$ values, necessitating the bounds $\delta \le p_j \le 1-\delta$ in \eqref{eq:p-bound}, but the variance for Li's Bayes estimator remains unaffected and does not need this extra restriction. On the other hand, Li's estimator \eqref{eq:li} is prone to bias if the missingness mechanism $p_{X}$ and the parameter for observed outcomes $\theta_X$ are correlated, \textit{i.e.}, if the missingness is missing-at-random (MAR) and not missing-completely-at-random (MCAR). We shall formalize this via the derived approximation for the mean squared error of the Bayes estimator in our theoretical properties section. \citet{linero2024nonparametric} argues that the `seemingly innocuous' priors like the one used by \citet{li2010bayesian} encode what is effectively a-priori knowledge that the amount of selection bias is minimal. 

% Note that $R_i \sim \text{Bernoulli}(p_{x_i})$ and under the assumption \eqref{eq:p-bound}, $\tilde{p} = 1/n \sum_i R_i$ acts as replacement for individual fixed $p_j$'s, and we can rewrite Li's estimator as:
% \begin{equation}
%     \hat{\psi}_{Li} = \frac{1}{n}\frac{\sum_{i=1}^{n} R_i Y_i + \alpha_F}{\frac{1}{n}(\sum_{i=1}^{n} R_i + 2\alpha_F)} = \frac{1}{n}\frac{\sum_{i=1}^{n} R_i Y_i + \alpha_F}{\tilde{p} + 2\alpha_F/n} = \asymp \frac{1}{n} \frac{\sum_{i=1}^{n} R_i Y_i}{\tilde{p}} \label{eq:li}
% \end{equation}
% This solves the problem by reducing a huge number $B$ of $p_j$ to an estimate $\tilde{p}$ and in the special case $p_j \equiv p, \forall j$, $\tilde{p} \to p$ for large $n$ providing face-validity to this estimator. 

\subsubsection{Binning and Smoothing \citep{ghosh2015weak}}

\citet{ghosh2015weak} provides another estimator by reducing the dimension of $(p_1, \ldots, p_B)$ by clubbing them into $k$ ($k \ll B$) groups by utilizing the boundedness assumption \eqref{eq:p-bound}. 
Based on this idea, we define the general binned-smoothed estimator as follows. 
\begin{definition}
Given a fixed $\delta \in (0,1)$ satisfying \eqref{eq:p-bound}, we divide the whole range $(\delta, 1-\delta) \ni p_b, b = 1, \ldots, B$ into $k$ sub-intervals $\{\delta_0 = \delta, \delta_1, \ldots, \delta_k = 1-\delta\}$ such that $\delta_{i+1}-\delta_i = (1-2\delta)/(k-1)$. We order the observed $p_{X_i}$'s into increasing order and define $\tilde{p}_j$ to be the mean of $p_{X_i}$ values in the $j^{th}$ partition consisting of $n_j$ points, i.e. 
\[
\tilde{p}_j = \begin{cases}
              &\frac{1}{n_j} \sum_{i: p_{X_i} \in (\delta_{j}, \delta_{j+1})} p_{X_i}, \; \text{if} \; n_j > 0 \\
              & (\delta_j + \delta_{j+1})/2 \; \text{otherwise}
              \end{cases} 
\]
Then, the \textit{binned-smoothed} estimator, based on \citep{ghosh2015weak} is:
\begin{equation}
    \hat{\psi}_{BS:HT} = \sum_{j=1}^{k} \frac{n_j}{n} \frac{1}{n_j} \frac{(\sum_{i=1}^{n_j} R_i Y_i)}{\tilde{p}_j}. \; \label{eq:ghosh}
\end{equation}
where $n_j$ is the number of $p_{X_i}$'s falling into the $j^{th}$ class. Note that, if the sample size $n \gg k$, the number of partitions, the number of $p_{X_i}$'s in any interval, i.e. $n_j$ values would typically be non-zero, and in the unlikely case $n_j = 0$ for any $j = 1, \ldots, k$, the corresponding term will not contribute to the numerator in \eqref{eq:ghosh}.
\end{definition}

\citet{ghosh2015weak}'s estimator performs at least as well as the HT estimator in small simulation studies. The idea of `binning and smoothing' can be applied to other inverse probability weighted estimators such as H\'ajek \eqref{eq:hajek}. The binned-smoothed H\'ajek estimator would take the form:
\begin{equation}
    \hat{\psi}_{BS:Hajek} = \frac{\sum_{j=1}^{k}\{(\sum_{i=1}^{n_j} R_i Y_i)/\tilde{p}_j\}}{\sum_{j=1}^{k}\{(\sum_{i=1}^{n_j} R_i)/\tilde{p}_j\}}, \; \label{eq:bs}
\end{equation}
where $n_j$ and $\tilde{p}_j$'s are as used in \eqref{eq:ghosh} before. 

\subsubsection{Bayesian Sieve \citep{sims2012example}}

We note that the idea of grouping the probabilities attached to the missingness indicators $R_i$'s, \textit{i.e.} clubbing $p_{X_i}$'s into $p_j$'s were also proposed in \citep{sims2012example}.
\citet{sims2012example} argues that estimators better than the HT estimator \eqref{eq:psi_ht} can be constructed using Bayesian approach if one acknowledges the existence of infinite dimensional unknown parameter in the model, e.g. assuming $p(b): 1, \ldots, B \mapsto (0,1)$ is known but not the conditional distribution of $[\theta \mid p]$. Sims suggested to break up the range of the known $p(\cdot)$ into $k$ segments, such that $\theta(b)$ and $p(b)$ are independent in each segment, and estimate the unknown $\theta(b)$ via a step function, with a constant for each segment. To complete Sim's Bayesian sieve, one needs to estimate the probability distribution on $p_j$'s induced by the partition, which will converge to a Dirichlet distribution for a large sample, making analytical calculation for the posterior mean of $\psi$ feasible.

\subsubsection{Gaussian Likelihood \citep{harmeling2007bayesian}}

\citet{harmeling2007bayesian} consider a slightly modified version of the Wasserman's model \eqref{eq:was}. Instead of a Bernoulli, they consider: 
\[
X_i \sim \UnifRV\{1, \ldots, B\}, \; Y_i \mid \Theta_{X_i} \sim \NormRV(\theta_{X_i}, 1), \; 1 \le i \le n, 
\]
and $\NormRV(\mu, 1)$ prior for each $\theta_b$, $b = 1, \ldots, B$, and a $\NormRV_{\mu}(0, \sigma)$ hyper-prior on $\mu$. The maximum likelihood estimate and the posterior mean Bayes' estimates for $B \to \infty$ are given by 
\begin{align*}
\hat{\psi}_{\text{MLE}} & = (\sum_{i=1}^n R_i)^{-1}(\sum_{i=1}^n R_i Y_i), \quad \text{and} \\
\hat{\psi}_{\text{Bayes}} & = (2/\sigma + \sum_{i=1}^n R_i)^{-1}(\sum_{i=1}^n R_i Y_i),
\end{align*}
respectively. Although these estimators are not directly comparable to $\eqref{eq:li}$, \citet{harmeling2007bayesian}'s likelihood-based estimators also achieve a lower MSE compared to the HT estimator, as expected. 

We also note that the idea of binning and smoothing $p_j$'s is also connected to importance sampling ideas in subsection \ref{sec:mc}, in particular, the trapezoidal rule (also called weighted Monte Carlo) by \citet{yakowitz1978weighted}, or its generalizations, called the Riemann summation approach \citet{philippe1997processing, philippe2001riemann}. These trapezoidal rules based on ordered samples reduces the variance drastically and achieves faster convergence and better stability. Along these lines, a later development, called the nested sampling approach by \citet{skilling2006nested} also relies on ``dividing the unit prior mass into tiny elements, and sorting them by likelihood." For a comprehensive discussion of the different strategies for variance reduction for the evidence estimation problem, we refer the readers to \citep{polson2014vertical, datta2023quantile}. 

\subsection{Properties of Li's Bayesian estimator}\label{sec:theory}

While a closed form analytical expression for the variance of the Li's estimator \eqref{eq:li} is difficult, we can use the linearization strategy aka the delta theorem to derive approximation for the mean and variance of the Bayes estimator \eqref{eq:li}. Doing so, we provide a proof and a closer look at an assertion in \citet{ritov2014bayesian} that the Li's estimator is consistent only if $\E(Y \mid R = 1) = \E(Y)$, and illustrate this phenomenon via simulation studies. 

Recall that for a pair of random variables $X,Y$ with mean $\mu_X$ and $\mu_Y$ and variances $\V(X), \V(Y)$ and covariance $\Cov(X,Y)$, we have the following approximations:
\begin{align}
\E(X/Y) & \approx \mu_X/\mu_Y \label{eq:delta-1}\\
\V(X/Y) & \approx (\frac{\mu_X}{\mu_Y}) \left(\frac{\V(X)}{\mu_X^2} + \frac{\V(Y)}{\mu_Y^2} - 2 \frac{\Cov(X,Y)}{\mu_X \mu_Y} \right)\label{eq:delta-2}
\end{align}

To calculate the approximate mean and variance for the Li's posterior mean estimator \eqref{eq:li}, we first calculate the mean and variances for the numerator and denominator separately as follows. Recall once again that the posterior mean estimator is given by $\hat{\psi}_{Li} = (\sum_i R_i Y_i + 1)/(\sum_i R_i + 2)$ assuming $\alpha_F = 1$, i.e., a $\text{Beta}(1,1)$ or Uniform prior on $\psi$ but the exact choice of shape parameter $\alpha_F$ does not influence the asymptotic nature of the variance. For notational convenience, we define the following quantities:
\begin{gather*}
\bar{\theta} = \frac{1}{B}\sum_{b=1}^{B} \theta_b\doteq \psi, \quad \bar{p} = \frac{1}{B}\sum_{b=1}^{B} p_b, \quad \bar{\theta.p} = \frac{1}{B}\sum_{b=1}^{B} \theta_b p_b, \\
\sigma_{\theta, p} =  \bar{\theta.p} - \bar{p} \bar{\theta} = \bar{\theta.p} - \bar{p} \psi. 
\end{gather*}
First, the expectation for the numerator and denominator follows from applying iterated expectations as:
\begin{align}
    \E(\sum_i R_i Y_i + 1) &= \E_X\{\sum_{i=1}^{n} \E(R_i \mid X_i) \E(Y_i \mid X_i)\} + 1  \nonumber\\
    & = \E_X\{ \sum_{i=1}^{n} \theta_{X_i} p_{X_i} \} + 1 = n \bar{\theta.p} + 1 =  n \psi \bar{p} + n \sigma_{\theta, p}+ 1 \; \label{eq:ex} \\
\E(\sum_i R_i + 2) & =  n \bar{p} + 2.  \label{eq:ey} 
\end{align}
Hence, the expectation for the Bayes' estimator can be approximated by taking the ratio of right hand sides from \eqref{eq:ex} and \eqref{eq:ey} as:
% \begin{multline*}
\[
   \E(\hat{\psi}_{Li}) \approx \frac{n \psi \bar{p} + n \sigma_{\theta, p} + 1}{n \bar{p} + 2} \to \psi + \sigma_{\theta, p}/\bar{p},\; \text{as} \; n \to \infty, 
   \]
% \end{multline*}
showing the $\hat{\psi}_{Li}$ is asymptotically unbiased if $\sigma_{\theta, p} = 0$, i.e., if $\theta_X \perp p_X \mid X$, or equivalently if $\E(Y \mid R = 1) = \E(Y)$, as claimed by \citep{ritov2014bayesian}. Now, the variances for the numerator and denominator can be calculated using the conditional variance identity. The expressions are given below. 
\begin{align}
    \V(\sum_i R_i + 2) & = \sum_i\V(R_i) \nonumber \\
    & = \sum_i \E [\V(R_i \mid X_i))] + \V [\E(R_i \mid X_i)] \nonumber \\
    & = \sum_i[ \E(p_{x_i}(1-p_{X_i}) + \V(p_{X_i})] = n(\bar{p} - \bar{p}^2). \label{eq:vy}
\end{align}
Similarly,
\begin{align}
    \V(\sum_i R_i Y_i + 1) & = \sum_i\V(R_i Y_i) =  n(\bar{\theta.p} - \bar{\theta.p}^2), \nonumber \\
    \quad \text{where} \; \bar{\theta.p} & = \frac{1}{B}\sum_{b=1}^{B} \theta_b p_b. \label{eq:vx}
\end{align}
Finally, the covariance term is:
\begin{align}
    \Cov\left(\sum_i R_i Y_i + 1,\; \sum_i R_i + 2\right) 
    &= \Cov\left(\sum_i R_i Y_i,\; \sum_i R_i\right) \nonumber \\
    &= \sum_i \Cov(R_i Y_i,\; R_i) \nonumber \\
    &= n \bar{\theta.p}~(1 - \bar{p}). \label{eq:cov}
\end{align}
% \begin{align}
%     \Cov(\sum_i R_i Y_i + 1 &, \sum_i R_i + 2) = \Cov(\sum_i R_i Y_i, \sum_i R_i) \nonumber \\
%     & = \E(\sum_i R_i^2 Y_i) - n \bar{p} \times (\psi n \bar{p} +n \sigma_{\theta, p}) \\
%     & = (\psi n \bar{p}+ n \sigma_{\theta, p}) (1- n \bar{p}). \; (R_i^2 \iidd R_i) \label{eq:cov}
% \end{align}

Putting everything together from \eqref{eq:ex}, \eqref{eq:ey}, \eqref{eq:vx}, \eqref{eq:vy} and \eqref{eq:cov}, we get the following formula for approximate variance for the Bayes estimator as:
\begin{multline}
\V(\hat{\psi}_{Li}) \approx 
\left(\frac{n \bar{\theta.p} + 1}{n \bar{p} + 2}\right)^2 \times \\ 
\left[ \frac{n(\bar{\theta.p} - \bar{\theta.p}^2)}{(n \bar{\theta.p} + 1)^2}
+ \frac{n(\bar{p} - \bar{p}^2)}{(n \bar{p} + 2)^2}
- \frac{2n \bar{\theta.p}(1 - \bar{p})}{(n \bar{p} + 2)(n \bar{\theta.p} + 1)} \right].
\label{eq:bayes-var}
\end{multline}

% \begin{multline}
%       \V(\hat{\psi}_{Li}) \approx \left(\frac{n \psi \bar{p} + n \sigma_{\theta, p} + 1}{n \bar{p} + 2}\right)^2 \times \\ \left[ \frac{n(\bar{\theta.p} - \bar{\theta.p}^2)}{(n \psi \bar{p} + 1)^2} + \frac{n(\bar{p} - \bar{p}^2)}{(n \bar{p} + 2)^2} - \frac{2n(\psi \bar{p} + \sigma_{\theta, p})(1- n \bar{p})}{(n \bar{p} + 2)\times (n \psi \bar{p} + 1)} \right]. \label{eq:bayes-var}  
% \end{multline}

A couple of immediate implications of the formula in \eqref{eq:bayes-var} are as follows. 
\begin{enumerate}
    \item Li's Bayesian estimator is consistent if $\theta_{X_i}$ and $p_{X_i}$ are uncorrelated, as $\E(\hat{\psi}_{Li}) \to \psi$ and $\V(\hat{\psi}_{Li}) = O(1/n) \to 0$ as the sample size $n \to \infty$. 
    \item Unlike the variance of the HT estimator, given in \eqref{eq:v-ht}, the variance of the Li's Bayes estimator does not have the individual $p_b$ terms in the denominator and will not inflate for very small $p_b$ values. In other words, the restriction \eqref{eq:p-bound} is not needed for the Bayesian solution. 
\end{enumerate}

\section{Simulation Examples}\label{sec:sim}
\subsection{Comparing IPW estimators for Wasserman's example} 
\subsubsection{Missing completely at random}
We compare the estimation performance for four different candidate estimators for the Robins-Ritov-Wasserman's problem. The candidates are: the original Horvitz--Thompson \eqref{eq:ht}, H\'ajek \eqref{eq:hajek}, the Li's estimator, \textit{i.e.} the Bayes posterior mean under a Beta hyperprior \eqref{eq:li} and \citet{ghosh2015weak}'s binning and smoothing idea applied to the H\'ajek estimator. We choose the bounds for $p_i$'s $\delta = 0.01$, and parameter space dimension $B = 1000$, \textit{i.e.} the $p_j$'s in our experiment are $B$ equidistant grid-points in $[\delta, 1-\delta]$. We vary the support of generative distribution for $\theta \in \UnifRV[a,b]$ to four different values, viz.  $[a,b] \in \{ [0.1,0.9], [0.1, 0.4], [0.35, 0.65], [0.6, 0.9]\}$. Finally, we take  sample size $n = 100$, as was originally intended in the RRW example to make it analogous to a high-dimensional problem with $B \gg n$.  

\begin{figure}[h!]
    \centering
    \begin{subfigure}{0.48\columnwidth}
        \includegraphics[width = \columnwidth]{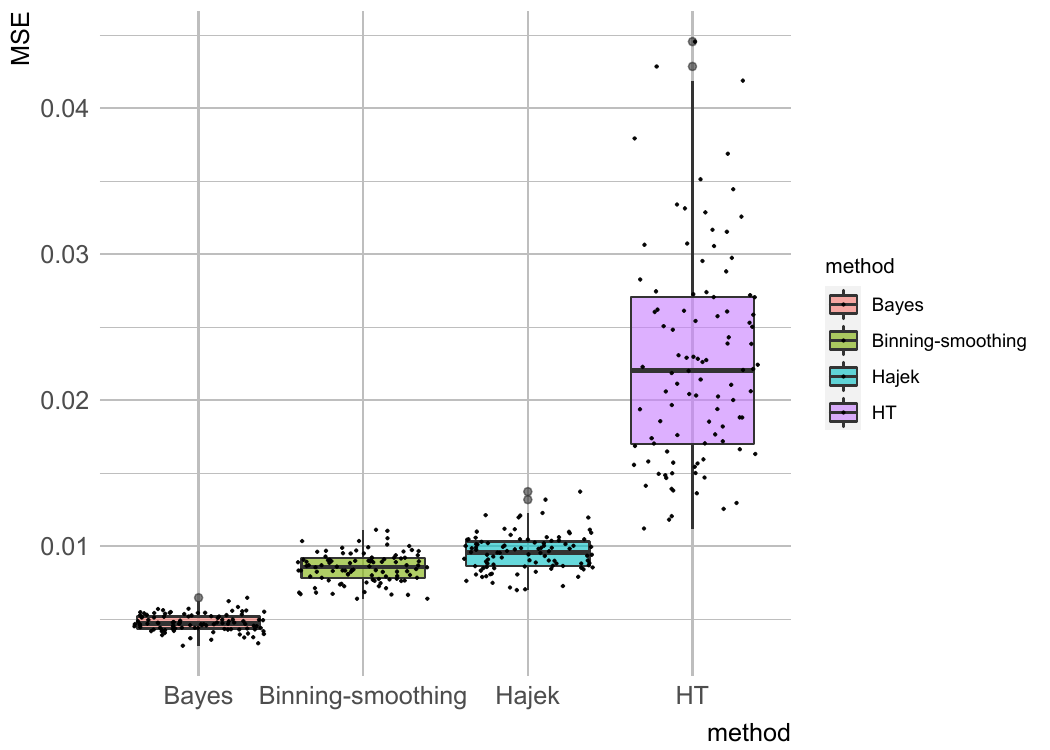}
        \caption{$\theta_b \sim \UnifRV(0.1, 0.9)$}
    \label{fig:2a}
    \end{subfigure}
    \begin{subfigure}{0.48\columnwidth}
        \includegraphics[width = \columnwidth]{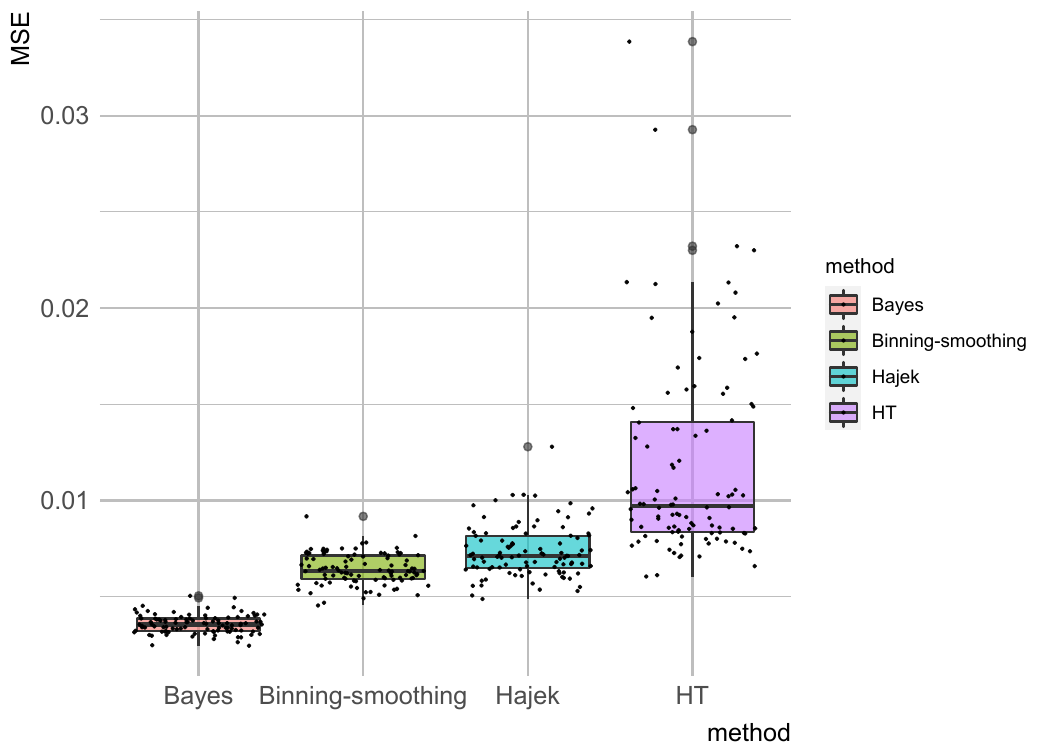}
        \caption{$\theta_b \sim \UnifRV(0.1, 0.4)$}
    \label{fig:2b}
    \end{subfigure}
        \begin{subfigure}{0.48\columnwidth}
        \includegraphics[width = \columnwidth]{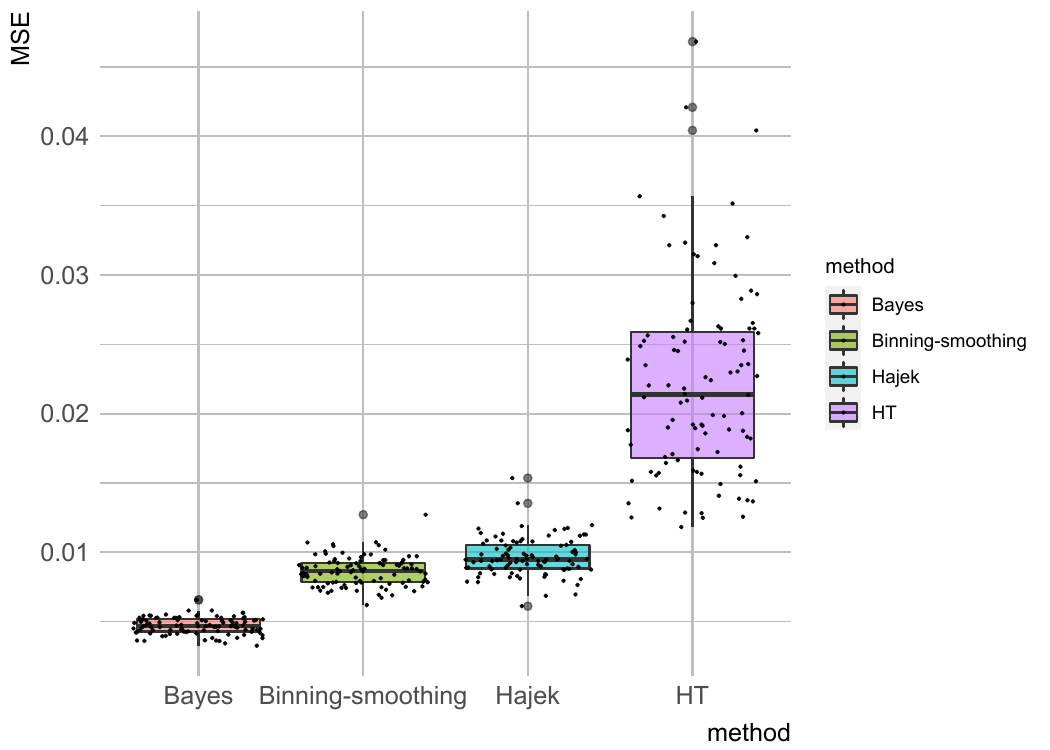}
        \caption{$\theta_b \sim \UnifRV(0.35, 0.65)$}
    \label{fig:2c}
    \end{subfigure}
        \begin{subfigure}{0.48\columnwidth}
        \includegraphics[width = \columnwidth]{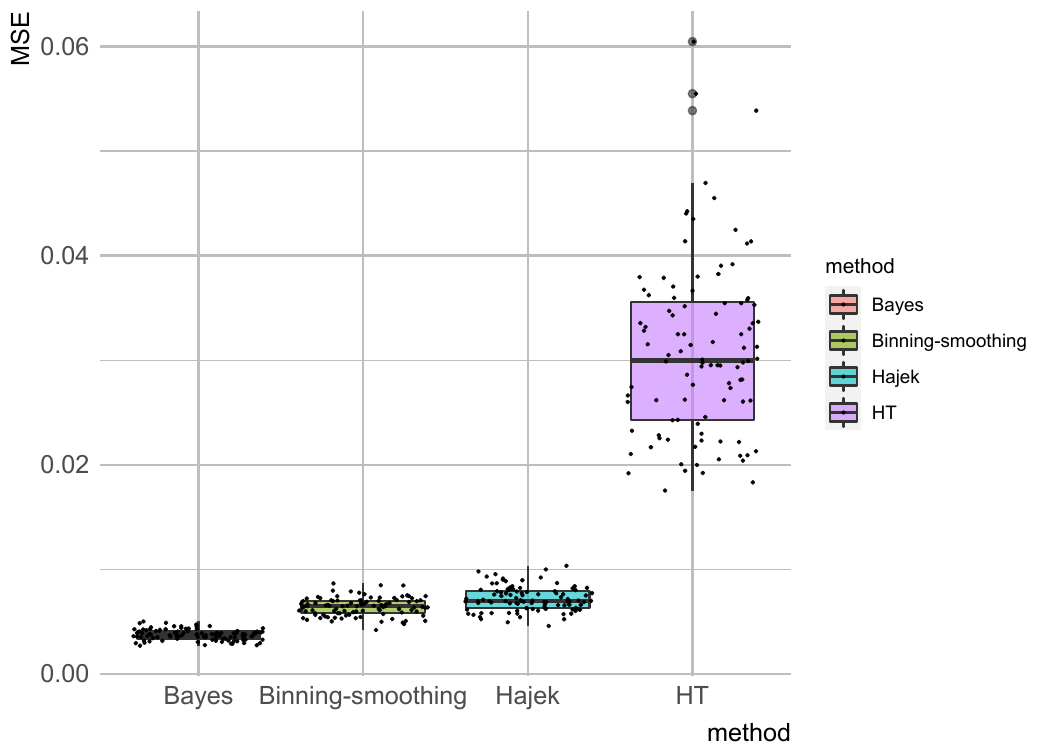}
        \caption{$\theta_b \sim \UnifRV(0.6, 0.49)$}
    \label{fig:2d}
    \end{subfigure}
    \caption{Mean square error comparison for Horvitz-Thompson, H\'ajek and the Li's estimator \eqref{eq:li} for different generative distributions of $\theta_1, \ldots, \theta_B$, and fixed values of $n, B$ and $\delta$. In each case, Bayes' estimator has the lowest MSE, followed by H\'ajek, and HT has the largest MSE.}
    \label{fig:2}
\end{figure}

Table \ref{tab:mse_comp} and Fig. \ref{fig:2} shows the mean squared errors calculated over $100$ replicates, and shows the following: (1) the Li's estimator leads to the lowest mean squared error over replicates, (2) the Horvitz--Thompson estimator performs the worst across all situations considered and finally, (3) the H\'ajek estimator and the Binning-Smoothing estimator \citep{ghosh2015weak} achieves very similar performance and generally occupies a middle ground between the Bayes' and the Horvitz--Thompson in terms of achieved mean-squared error. 

\begin{table}[ht]
 \caption{Mean squared error comparison between the four candiate estimators: Horvitz--Thompson \eqref{eq:ht}, H\'ajek \eqref{eq:hajek}, the Li's Bayesian estimator \eqref{eq:li} and \citet{ghosh2015weak}'s estimator for different values of $[a,b]$, where $\theta \in [a,b]$ . The numbers on the table are reported after multiplying the MSE by $10^2$ for better comparison and the column winner are denoted by boldfaced entries.}
   \centering
  \footnotesize{
    \begin{tabular}{l|crrr}
    \toprule
          & \multicolumn{2}{c}{[0.6, 0.9]} & \multicolumn{2}{c}{[0.1, 0.9]} \\
    \midrule
    method & mean  & sd    & mean  & sd \\
    Bayes' (Li's) & \textbf{0.37198} & 0.05093 & \textbf{0.47541} & 0.05980 \\
    Binning-smoothing & 0.63991 & 0.08583 & 0.85225 & 0.10204 \\
    H\'ajek & 0.71787 & 0.11643 & 0.95824 & 0.12818 \\
    HT    & 3.09007 & 0.83348 & 2.27093 & 0.72187 \\
    \bottomrule
          &       &       &       &  \\
          & \multicolumn{2}{c}{[0.1, 0.4]} & \multicolumn{2}{c}{[0.35, 0.65]} \\
  \toprule
    method & mean  & sd    & mean  & sd \\
    % \midrule
    Bayes' (Li's) & \textbf{0.35743} & 0.04758 & \textbf{0.47329} & 0.06272 \\
    Binning-smoothing & 0.64379 & 0.08114 & 0.86459 & 0.10523 \\
    H\'ajek & 0.73526 & 0.13494 & 0.96710 & 0.13523 \\
    HT    & 1.17543 & 0.51219 & 2.22213 & 0.69317 \\
    \bottomrule
    \end{tabular}%
    }
  \label{tab:mse_comp}%
\end{table}%

\begin{figure}[ht!]
    \centering
    \begin{subfigure}{0.45\columnwidth}
        \includegraphics[width = \columnwidth]{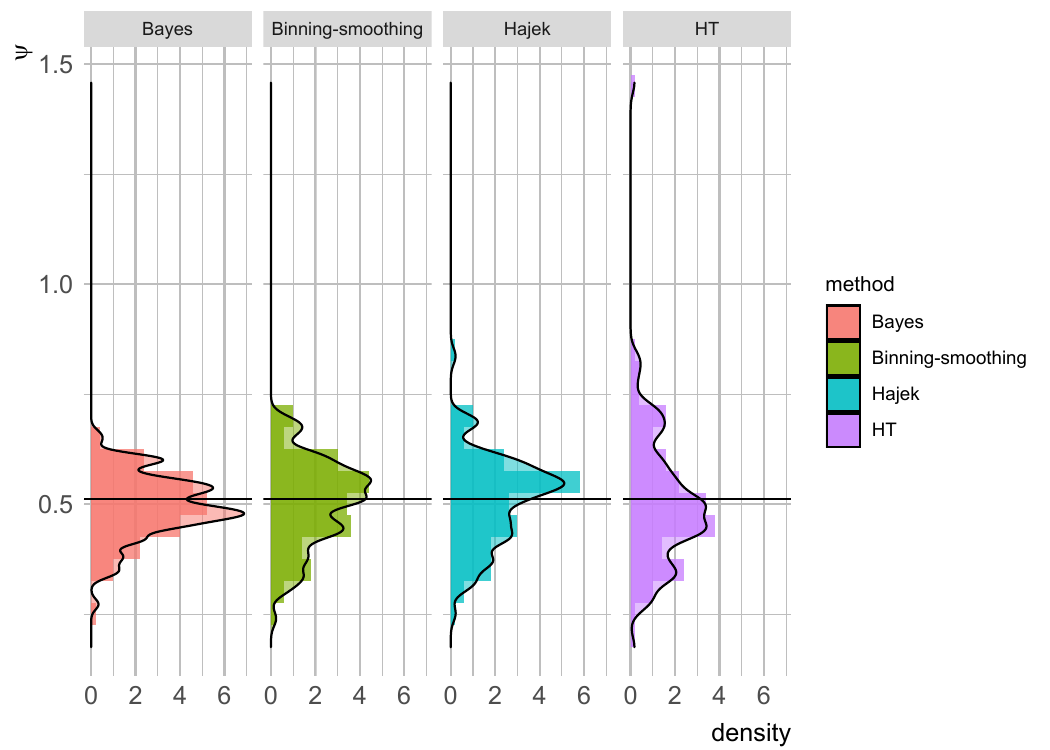}
        \caption{$\theta_b \sim \UnifRV(0.1, 0.9)$}
    \label{fig:1a}
    \end{subfigure}
    \begin{subfigure}{0.45\columnwidth}
        \includegraphics[width = \columnwidth]{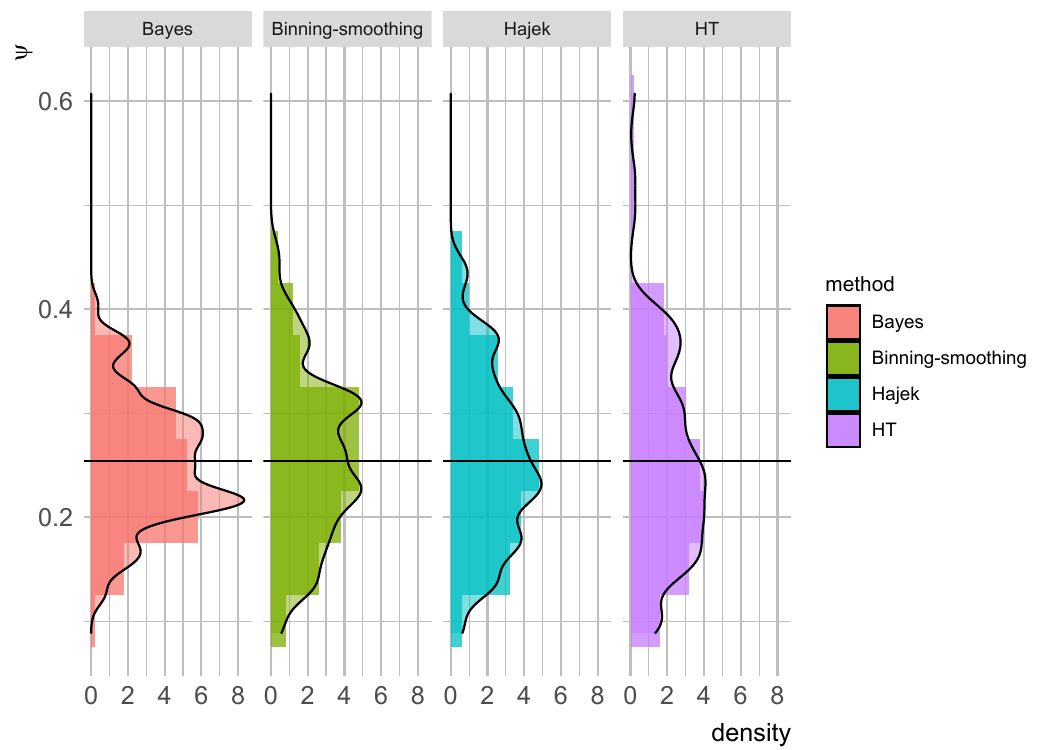}
        \caption{$\theta_b \sim \UnifRV(0.1, 0.4)$}
    \label{fig:1b}
    \end{subfigure}
        \begin{subfigure}{0.45\columnwidth}
        \includegraphics[width = \columnwidth]{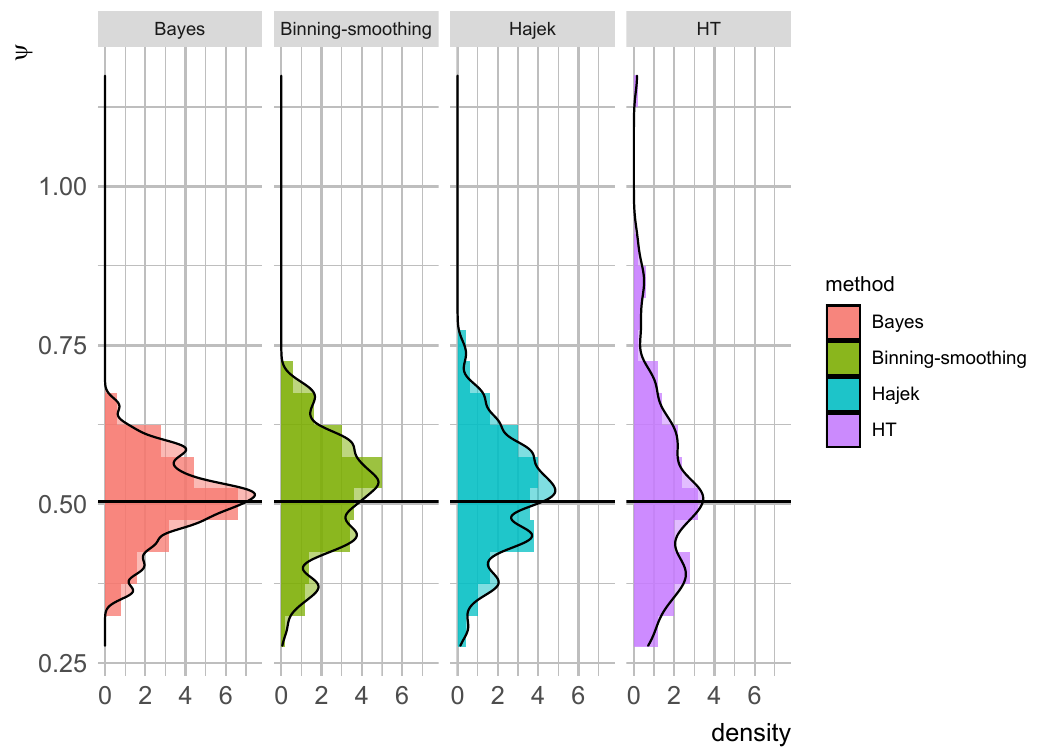}
        \caption{$\theta_b \sim \UnifRV(0.35, 0.65)$}
    \label{fig:1c}
    \end{subfigure}
        \begin{subfigure}{0.45\columnwidth}
        \includegraphics[width = \columnwidth]{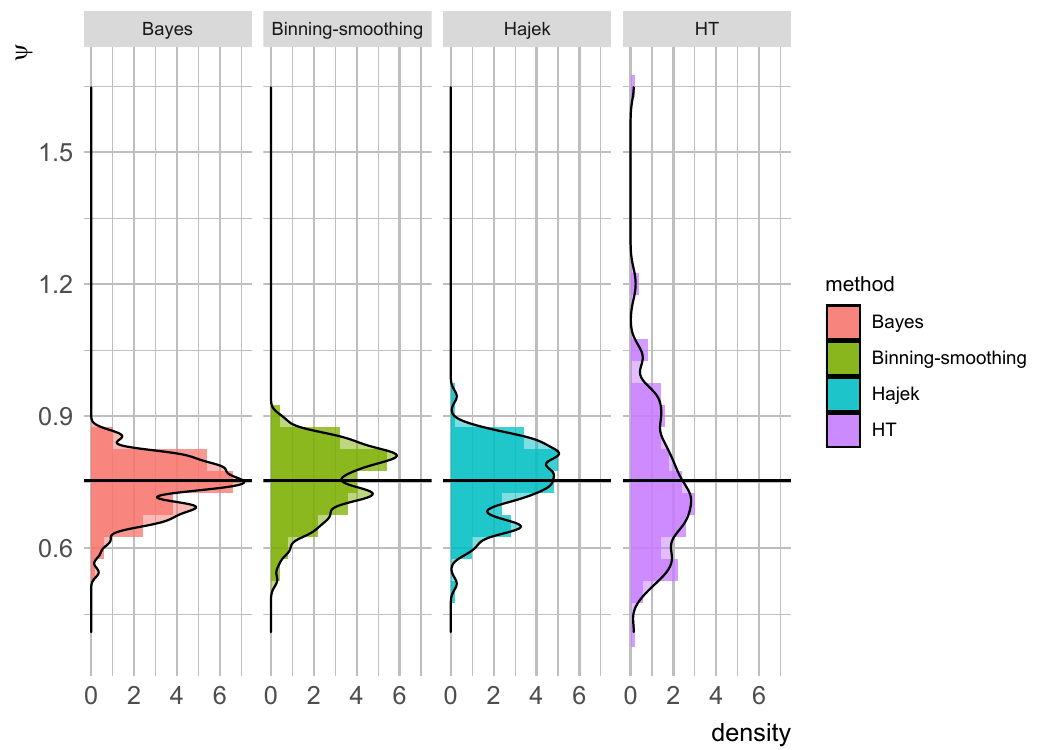}
        \caption{$\theta_b \sim \UnifRV(0.6, 0.49)$}
    \label{fig:1d}
    \end{subfigure}
    \caption{Histogram and Kernel Density plots showing the distribution of Horvitz--Thompson, H\'ajek and the Bayes estimators for different generative distributions of $\theta_1, \ldots, \theta_B$, and fixed values of $n, B$ and $\delta$. }
    \label{fig:1}
\end{figure}

For a single replication with $1,000$ evaluations, Fig. \ref{fig:1} shows the bias and variance for the four candidate estimators. Fig. \ref{fig:1} demonstrates the nature of variance reduction by Bayes' estimator and the Binning-smoothing idea in estimating $\psi = \E(\btheta)$ for different ranges of $\theta$, without any significant increase in bias. 

\subsubsection{Missing at random} As discussed earlier, Li's estimator can be biased when there is a non-zero correlation between $\theta_X$ and $p_X$, \textit{e.g.} when the missingness might arise due to confounding. We show an example here to illustrate the effect of this situation, and compare the four candidates. We take $\delta = 0.1$, \textit{i.e.} $p_{b}$'s to be equidistant points in $[0.1, 0.9]$, and instead of $\theta_b$'s to be uniformly distributed in $[a,b]$, we take:
\begin{align*}
    \theta_b & = \gamma \times p_b + (1-\gamma) \times U_b, \; \\ \text{where} &\; U_b \sim \UnifRV(a,b), \; \gamma = 0.5, b = 1, \ldots, B.
\end{align*}

\begin{figure}[h!]
    \centering
    \begin{subfigure}{0.49\columnwidth}
        \includegraphics[width = \columnwidth]{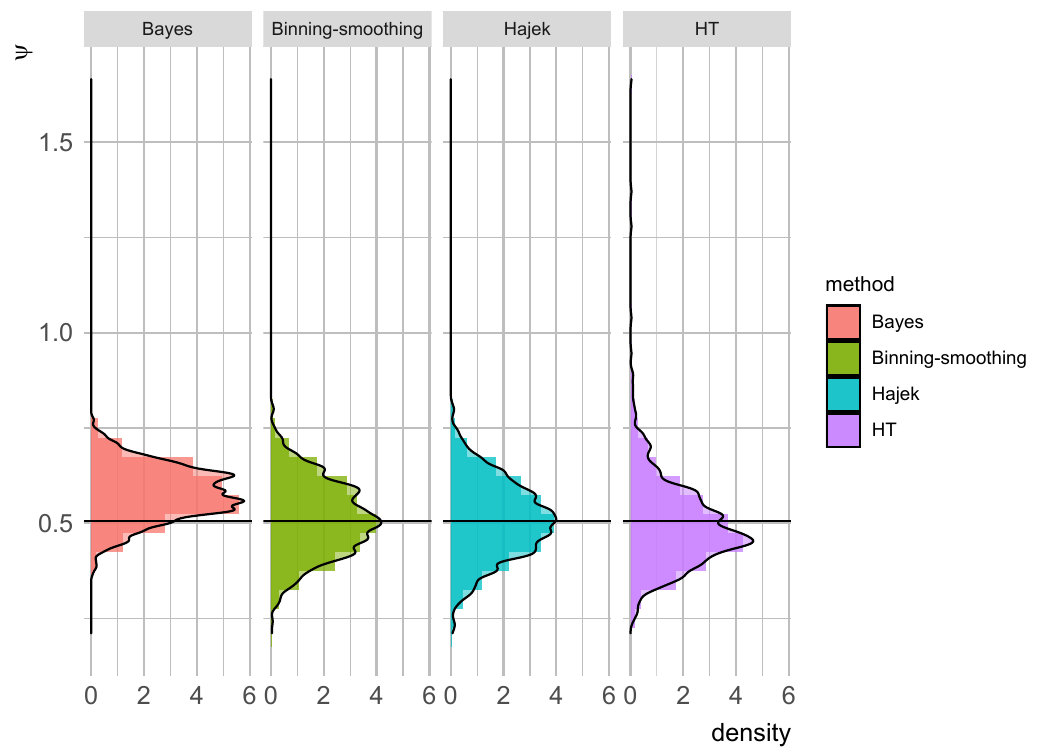}
        \caption{$\theta_b = \gamma \times p_b + (1-\gamma) \times \UnifRV(0,1)$}
    \label{fig:4a}
    \end{subfigure}
    \begin{subfigure}{0.49\columnwidth}
        \includegraphics[width = \columnwidth]{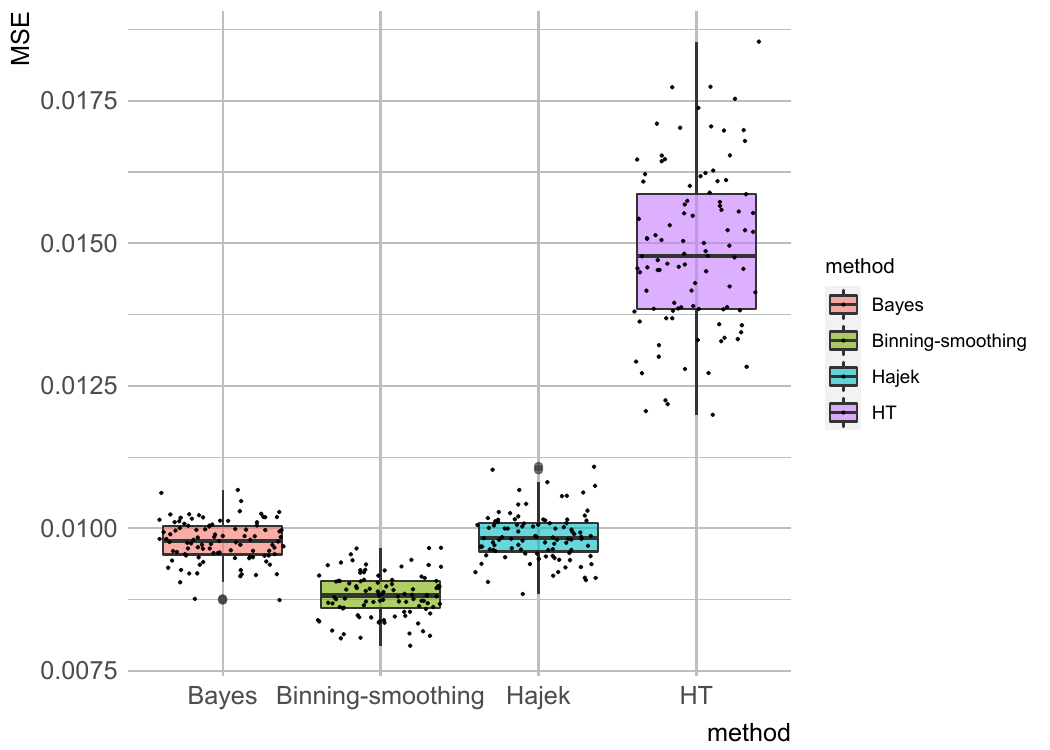}
        \caption{$\theta_b = \gamma \times p_b + (1-\gamma) \times \UnifRV(0,1)$}
    \label{fig:4b}
    \end{subfigure}
    \caption{Histogram, Kernel density plots and MSE comparison for Horvitz--Thompson, H\'ajek and the Bayes estimators for the missing-at-random scenario. }
    \label{fig:4}
\end{figure}

The remaining set-up is same as before. Clearly, in this case the true value of $\psi = \E(\theta) = 0.5$. Figure \ref{fig:4a} shows that the Li's estimator has an upward bias, as expected, because of the positive correlation between $\theta$ and $p$, but the H\'ajek and Binned-Smoothed estimators are unaffected. On the other hand, the Binned-Smoothed is significantly better than all the other candidates (HT, H\'ajek and Li's) in terms of MSE (Fig. \ref{fig:4b}). This shows that the Li's posterior mean estimator can be biased when there is confounding and one needs to be careful when using it. The binned-smoothed method, on the other hand, performs well irrespective of the correlation between $\theta_X$ and $p_X$, \textit{i.e.} for both MAR and MCAR-type missingness. 

\section{Discussion}\label{sec:dis}

Our main goal in this paper was to shed light on various aspects of the family of inverse probability weight estimators, discuss its weakness and strengths and highlight the connections between survey sampling and Monte Carlo integration via these popular tool pervasive in Statistical literature. For survey sampling, we consider some of the `weak paradoxes' \citep{ghosh2015weak} that arise from using inverse probability estimators using the well-known examples from \citet{wasserman2004bayesian} and \citet{basu1988statistical}, and review the merits and demerits of popular IPW estimators. In particular, we show that the hierarchical Bayes' estimator due to \citep{li2010bayesian} leads to robustness against pathological situations but admits bias in presence of confounding. We provide sufficient conditions for consistency of the Li's posterior mean estimate in Wasserman's example and show that, under certain conditions, both the Li's estimator and the binning-smoothing idea \citep{ghosh2015weak} achieves lower variation in mean squared errors. We then highlight the analogous tools in Monte Carlo integration, revisiting a few earlier works \citep[e.g.][]{hesterberg1988advances}, where the Horvitz--Thompson estimator is likened to the na\"ive importance sampling, and much like survey sampling, self-normalized weights or using control variates lead to better estimators. We conclude with a brief discussion of applications in the context of causal inference, a key use of IPW estimators, and a few possible directions for future work. 

A possible future direction, borrowing from \citep{kong2003theory} is to use the general semiparametric models and the associated estimators or computational algorithm for $k > 1$ in the context of survey sampling that would allow one to combine data from one or more surveys that use different but known inclusion probabilities. We conjecture that this might lead to improved estimators for average treatment effect estimation, which we briefly describe next.

\noindent \textbf{Average Treatment Effect Estimation:} A natural application of IPW estimator is average treatment effect (ATE) estimation in potential outcomes causal inference framework as pointed out in \citep{delevoye2020consistency, khan2023adaptive}. We refer the readers to the excellent references in \citep{cunningham2021causal, rubin1974estimating, imbens2004nonparametric} for in-depth coverage and historical backgrounds. Here we measure the difference in potential outcomes observed over time points $t_2 > t_1$, where only one of the two potential outcomes are observed for any unit. Using \citet{khan2023adaptive}'s notation: we have the triplets $(Y_k(0), Y_k(1), p_k)$ for $k = 1, \ldots, n$, where we observe $Y_k(I_k)$ for $I_k \sim \text{Bernoulli}(p_k)$. and the parameter of interest is $ATE = \tau = \E[Y_k(1) - Y_k(0)]$ from the observations available to us. The IPW estimators are employed here to estimate the two population means $\psi_1 = \E[Y_k(1)]$ and $\psi_0 = \E[Y_k(0)]$ separately and estimating $\tau$ by $\hat{\tau} = \hat{\psi_1} - \hat{\psi_0}$. Estimating the population means $\psi_j$, $j = 1,2$ turns ATE into a survey sampling problem and one can use all the estimators: HT, H\'ajek or the various improvements such as the adaptive IPW estimator by \citet{khan2023adaptive} based on the Trotter--Tukey idea \citep{trotter1956conditional} here. \citet{khan2023adaptive} further develop an adaptive estimator by minimizing the variance of between-group differences and show that both the separate AIPW and the joint AIPW estimators achieve lower mean squared errors compared to the usual HT and H\'ajek.

Thus, a second possible direction for future research is comparing a suitable modification of the Bayes estimator in \eqref{eq:li} for the ATE problem with that of the adaptive estimators developed by \citet{khan2023adaptive} and both empirically and theoretically to investigate consistency properties. It will be also worthwhile to consider the semiparametric model in \citep{kong2003theory} for ATE and compare the results of simultaneous estimation using MLE with these candidates. 

Finally, the problems presented in \citep{wasserman2004bayesian} and discussed in \citep{sims2010understanding, sims2007thinking} are inherently high-dimensional in nature, where sparsity is pervasive. There is a large and growing literature on promoting sparsity in causal inference, \textit{e.g.} in presence of a large number of confounders or baseline variables while estimating the effect of an exposure on an outcome. We refer the readers to \citet{shortreed2017outcome, wang2012bayesian, wang2015accounting,kim2022bayesian} for recent advances in this area. 
In our simple focal example, the high-dimensional parameter $\btheta = (\theta_1, \ldots, \theta_B)$ or $\p = (p_1, \ldots, p_B)$ could exhibit sparsity with a small $\ell_0$ norm or other notions of sparsity. To handle sparsity in higher dimensions while maintaining tail-robustness and accuracy, the state-of-the-art Bayesian solution would be augmenting \eqref{eq:was} with global-local shrinkage priors \citep{polson2010shrink, polson2012local, bhadra2019lasso}, \textit{i.e.}
\begin{align*}
    \theta_b \mid \alpha_T, \psi & \ind f_\text{HS}(\theta; \alpha_T \psi, \; \alpha_T (1-\psi_0), \; \tau) \\
  f_{\text{HS}}(\theta ; a, b, \tau) & = \theta^{a-1}(1 - \theta)^{b-1}\{1-(1-\theta)\tau^2\}^{-(a+b)} , \\
    (\psi, \alpha_T \mid \alpha_F) & \sim \text{Beta}(\alpha_F, \alpha_F) \times \pi(\alpha_T), \\
        &\; \theta, \psi \in [0,1].
\end{align*}
Here, as before, $\psi$ is the mean of $\theta$, $\alpha_T$ and $\alpha_F$ are the and the shape parameters for $\btheta$ and $\psi$ respectively and $\tau$ is a global shrinkage parameter adjusting to sparsity. Such Bayesian regularization priors have been used successfully for treatment effect estimation with more control variates than observations \citep{hahn2018regularization}. Designing built-in sparsity priors that can also incorporate selection mechanism for confounder selection is an interesting problem that we plan to address in a future endeavor.

% \begin{acks}[Acknowledgments]
% We thank a referee for constructive comments on an earlier version of the manuscript. 
% \end{acks}
%
%\begin{funding}
%Dr. Datta acknowledges support from the National Science Foundation (DMS-2015460). 
%\end{funding}

% \end{acks}
\section*{Acknowledgement}
We thank a referee for constructive comments on an earlier version of the manuscript. 

\section*{Funding}
Dr. Datta acknowledges support from the National Science Foundation (DMS-2015460).

% \newpage
\vskip 0.2in
\bibliographystyle{ba}
\bibliography{references,hs-review}

\begin{thebibliography}{}

\bibitem[Basu, 1988]{basu1988statistical}
Basu, D. (1988).
\newblock {\em Statistical information and likelihood: a collection of critical
  essays by Dr. D. Basu (J. K. Ghosh, Ed.)}, volume~45.
\newblock Lecture Notes in Statistics, Springer Science \& Business Media.

\bibitem[Bhadra et~al., 2019]{bhadra2019lasso}
Bhadra, A., Datta, J., Polson, N.~G., and Willard, B. (2019).
\newblock Lasso meets horseshoe.
\newblock {\em Statistical Science}, 34(3):405--427.

\bibitem[Chauvet, 2014]{chauvet2014note}
Chauvet, G. (2014).
\newblock A note on the consistency of the narain-horvitz-thompson estimator.
\newblock {\em arXiv preprint arXiv:1412.2887}.

\bibitem[Chen et~al., 2017]{chen2017approaches}
Chen, Q., Elliott, M.~R., Haziza, D., Yang, Y., Ghosh, M., Little, R.~J.,
  Sedransk, J., and Thompson, M. (2017).
\newblock Approaches to improving survey-weighted estimates.
\newblock {\em Statistical Science}, 32(2):227--248.

\bibitem[Chopin and Robert, 2010]{chopin2010properties}
Chopin, N. and Robert, C.~P. (2010).
\newblock Properties of nested sampling.
\newblock {\em Biometrika}, 97(3):741--755.

\bibitem[Cunningham, 2021]{cunningham2021causal}
Cunningham, S. (2021).
\newblock Causal inference.
\newblock In {\em Causal Inference}. Yale University Press.

\bibitem[Datta and Polson, 2023]{datta2023quantile}
Datta, J. and Polson, N.~G. (2023).
\newblock Quantile importance sampling.
\newblock {\em arXiv preprint arXiv:2305.03158}.

\bibitem[Delevoye and S{\"a}vje, 2020]{delevoye2020consistency}
Delevoye, A. and S{\"a}vje, F. (2020).
\newblock Consistency of the horvitz--thompson estimator under general sampling
  and experimental designs.
\newblock {\em Journal of Statistical Planning and Inference}, 207:190--197.

\bibitem[Diaconis, 1988]{diaconis1988bayesian}
Diaconis, P. (1988).
\newblock Bayesian numerical analysis.
\newblock {\em Statistical decision theory and related topics IV}, 1:163--175.

\bibitem[Efron and Morris, 1977]{efron1977stein}
Efron, B. and Morris, C. (1977).
\newblock Stein's paradox in statistics.
\newblock {\em Scientific American}, 236(5):119--127.

\bibitem[Firth, 2011]{firth2011improved}
Firth, D. (2011).
\newblock On improved estimation for importance sampling.
\newblock {\em Brazilian Journal of Probability and Statistics},
  25(3):437--443.

\bibitem[Ghosh, 2015]{ghosh2015weak}
Ghosh, J.~K. (2015).
\newblock Weak paradoxes and paradigms.
\newblock In {\em Statistical Paradigms: Recent Advances and Reconciliations},
  pages 3--12. World Scientific.

\bibitem[Hahn et~al., 2018]{hahn2018regularization}
Hahn, P.~R., Carvalho, C.~M., Puelz, D., and He, J. (2018).
\newblock Regularization and confounding in linear regression for treatment
  effect estimation.
\newblock {\em Bayesian Analysis}, 13(1):163--182.

\bibitem[Harmeling and Touissant, 2007]{harmeling2007bayesian}
Harmeling, S. and Touissant, M. (2007).
\newblock Bayesian estimators for robins-ritov's problem.
\newblock Technical report.

\bibitem[Haziza and Beaumont, 2017]{haziza2017construction}
Haziza, D. and Beaumont, J.-F. (2017).
\newblock Construction of weights in surveys: A review.

\bibitem[Hesterberg, 1995]{hesterberg1995weighted}
Hesterberg, T. (1995).
\newblock Weighted average importance sampling and defensive mixture
  distributions.
\newblock {\em Technometrics}, 37(2):185--194.

\bibitem[Hesterberg, 1988]{hesterberg1988advances}
Hesterberg, T.~C. (1988).
\newblock {\em Advances in importance sampling}.
\newblock PhD thesis, Stanford University.

\bibitem[Horvitz and Thompson, 1952]{horvitz1952generalization}
Horvitz, D.~G. and Thompson, D.~J. (1952).
\newblock A generalization of sampling without replacement from a finite
  universe.
\newblock {\em Journal of the American statistical Association},
  47(260):663--685.

\bibitem[Hájek, 1971]{hajek1971comment}
Hájek, J. (1971).
\newblock Comment on ``an essay on the logical foundations of survey sampling''
  by basu.
\newblock In Godambe, V.~P. and Sprott, D.~A., editors, {\em Foundations of
  Statistical Inference}, pages 236--242. Holt, Rinehart and Winston.

\bibitem[Imbens, 2004]{imbens2004nonparametric}
Imbens, G.~W. (2004).
\newblock Nonparametric estimation of average treatment effects under
  exogeneity: A review.
\newblock {\em Review of Economics and statistics}, 86(1):4--29.

\bibitem[Isaki and Fuller, 1982]{isaki1982survey}
Isaki, C.~T. and Fuller, W.~A. (1982).
\newblock Survey design under the regression superpopulation model.
\newblock {\em Journal of the American Statistical Association},
  77(377):89--96.

\bibitem[James and Stein, 1961]{james1961estimation}
James, W. and Stein, C. (1961).
\newblock Estimation with quadratic loss.
\newblock In {\em Proceedings of the Fourth Berkeley Symposium on Mathematical
  Statistics and Probability, Volume 1: Contributions to the Theory of
  Statistics}, pages 361--379. University of California Press.

\bibitem[Khan and Ugander, 2023]{khan2023adaptive}
Khan, S. and Ugander, J. (2023).
\newblock Adaptive normalization for ipw estimation.
\newblock {\em Journal of Causal Inference}, 11(1):20220019.

\bibitem[Kim et~al., 2022]{kim2022bayesian}
Kim, C., Tec, M., and Zigler, C.~M. (2022).
\newblock Bayesian nonparametric adjustment of confounding.
\newblock {\em arXiv preprint arXiv:2203.11798}.

\bibitem[Kong et~al., 2003]{kong2003theory}
Kong, A., McCullagh, P., Meng, X.-L., Nicolae, D., and Tan, Z. (2003).
\newblock A theory of statistical models for monte carlo integration.
\newblock {\em Journal of the Royal Statistical Society: Series B (Statistical
  Methodology)}, 65(3):585--604.

\bibitem[Li, 2010]{li2010bayesian}
Li, L. (2010).
\newblock Are bayesian inferences weak for wasserman's example?
\newblock {\em Communications in Statistics—Simulation and
  Computation{\textregistered}}, 39(3):655--667.

\bibitem[Linero, 2024]{linero2024nonparametric}
Linero, A.~R. (2024).
\newblock In nonparametric and high-dimensional models, bayesian ignorability
  is an informative prior.
\newblock {\em Journal of the American Statistical Association},
  119(548):2785--2798.

\bibitem[Little, 2008]{little2008weighting}
Little, R.~J. (2008).
\newblock Weighting and prediction in sample surveys.
\newblock {\em Calcutta Statistical Association Bulletin}, 60(3-4):147--167.

\bibitem[Madrid-Padilla et~al., 2018]{madrid2018deconvolution}
Madrid-Padilla, O.-H., Polson, N.~G., and Scott, J. (2018).
\newblock A deconvolution path for mixtures.
\newblock {\em Electronic Journal of Statistics}, 12(1):1717--1751.

\bibitem[Narain, 1951]{narain1951sampling}
Narain, R. (1951).
\newblock On sampling without replacement with varying probabilities.
\newblock {\em Journal of the Indian Society of Agricultural Statistics},
  3(2):169--175.

\bibitem[Neyman, 1934]{neyman1934two}
Neyman, J. (1934).
\newblock On the two different aspects of the representative method: The method
  of stratified sampling and the method of purposive selection.
\newblock {\em Journal of the Royal Statistical Society Series A: Statistics in
  Society}, 97(4):558--606.

\bibitem[Owen, 2019]{owen2019comment}
Owen, A.~B. (2019).
\newblock Comment: Unreasonable effectiveness of monte carlo.
\newblock {\em Statistical Science}, 34(1):29--33.

\bibitem[Philippe, 1997]{philippe1997processing}
Philippe, A. (1997).
\newblock Processing simulation output by riemann sums.
\newblock {\em Journal of Statistical Computation and Simulation},
  59(4):295--314.

\bibitem[Philippe and Robert, 2001]{philippe2001riemann}
Philippe, A. and Robert, C.~P. (2001).
\newblock Riemann sums for mcmc estimation and convergence monitoring.
\newblock {\em Statistics and Computing}, 11(2):103--115.

\bibitem[Polson and Scott, 2010]{polson2010shrink}
Polson, N.~G. and Scott, J.~G. (2010).
\newblock Shrink globally, act locally: Sparse {{Bayesian}} regularization and
  prediction.
\newblock {\em Bayesian Statistics}, 9:501--538.

\bibitem[Polson and Scott, 2012]{polson2012local}
Polson, N.~G. and Scott, J.~G. (2012).
\newblock Local shrinkage rules, l{\'e}vy processes and regularized regression.
\newblock {\em Journal of the Royal Statistical Society: Series B (Statistical
  Methodology)}, 74(2):287--311.

\bibitem[Polson and Scott, 2014]{polson2014vertical}
Polson, N.~G. and Scott, J.~G. (2014).
\newblock Vertical-likelihood monte carlo.
\newblock {\em arXiv preprint arXiv:1409.3601}.

\bibitem[Raftery et~al., 2006]{raftery2006estimating}
Raftery, A.~E., Newton, M.~A., Satagopan, J.~M., and Krivitsky, P.~N. (2006).
\newblock Estimating the integrated likelihood via posterior simulation using
  the harmonic mean identity.

\bibitem[Ramakrishnan, 1973]{ramakrishnan1973alternative}
Ramakrishnan, M. (1973).
\newblock An alternative proof of the admissibility of the horvitz-thompson
  estimator.
\newblock {\em The Annals of Statistics}, 1(3):577--579.

\bibitem[Rao et~al., 1999]{rao1999some}
Rao, J.~N., Chaudhuri, A., Eltinge, J., Fay, R.~E., Ghosh, J., Ghosh, M.,
  Lahiri, P., and Pfeffermann, D. (1999).
\newblock Some current trends in sample survey theory and methods (with
  discussion).
\newblock {\em Sankhy{\=a}: The Indian Journal of Statistics, Series B}, pages
  1--57.

\bibitem[Ritov et~al., 2014]{ritov2014bayesian}
Ritov, Y., Bickel, P.~J., Gamst, A.~C., and Kleijn, B. J.~K. (2014).
\newblock The bayesian analysis of complex, high-dimensional models: Can it be
  coda?
\newblock {\em Statistical Science}, 29(4):619--639.

\bibitem[Robins and Ritov, 1997]{robins1997toward}
Robins, J.~M. and Ritov, Y. (1997).
\newblock Toward a curse of dimensionality appropriate (coda) asymptotic theory
  for semi-parametric models.
\newblock {\em Statistics in medicine}, 16(3):285--319.

\bibitem[Rubin, 1974]{rubin1974estimating}
Rubin, D.~B. (1974).
\newblock Estimating causal effects of treatments in randomized and
  nonrandomized studies.
\newblock {\em Journal of educational Psychology}, 66(5):688--701.

\bibitem[S{\"a}rndal et~al., 2003]{sarndal2003model}
S{\"a}rndal, C.-E., Swensson, B., and Wretman, J. (2003).
\newblock {\em Model assisted survey sampling}.
\newblock Springer Science \& Business Media.

\bibitem[Shortreed and Ertefaie, 2017]{shortreed2017outcome}
Shortreed, S.~M. and Ertefaie, A. (2017).
\newblock Outcome-adaptive lasso: variable selection for causal inference.
\newblock {\em Biometrics}, 73(4):1111--1122.

\bibitem[Sims, 2010]{sims2010understanding}
Sims, C. (2010).
\newblock Understanding non-bayesians.
\newblock {\em Unpublished chapter, Department of Economics, Princeton
  University}.

\bibitem[Sims, 2007]{sims2007thinking}
Sims, C.~A. (2007).
\newblock Thinking about instrumental variables.
\newblock {\em manuscript, Department of Economics, Princeton University}.

\bibitem[Sims, 2012]{sims2012example}
Sims, C.~A. (2012).
\newblock On an example of larry wasserman, round 2.

\bibitem[Skilling, 2006]{skilling2006nested}
Skilling, J. (2006).
\newblock Nested sampling for general bayesian computation.
\newblock {\em Bayesian analysis}, 1(4):833--859.

\bibitem[Stigler, 1990]{stigler19901988}
Stigler, S.~M. (1990).
\newblock The 1988 neyman memorial lecture: a galtonian perspective on
  shrinkage estimators.
\newblock {\em Statistical Science}, pages 147--155.

\bibitem[Trotter and Tukey, 1956]{trotter1956conditional}
Trotter, H.~F. and Tukey, H. (1956).
\newblock Conditional monte carlo for normal samples.
\newblock In {\em Proc. Symp. on Monte Carlo Methods}, pages 64--79. John Wiley
  and Sons.

\bibitem[Wang et~al., 2015]{wang2015accounting}
Wang, C., Dominici, F., Parmigiani, G., and Zigler, C.~M. (2015).
\newblock Accounting for uncertainty in confounder and effect modifier
  selection when estimating average causal effects in generalized linear
  models.
\newblock {\em Biometrics}, 71(3):654--665.

\bibitem[Wang et~al., 2012]{wang2012bayesian}
Wang, C., Parmigiani, G., and Dominici, F. (2012).
\newblock Bayesian effect estimation accounting for adjustment uncertainty.
\newblock {\em Biometrics}, 68(3):661--671.

\bibitem[Wasserman, 2004]{wasserman2004bayesian}
Wasserman, L. (2004).
\newblock Bayesian inference.
\newblock In {\em All of Statistics}, pages 175--192. Springer.

\bibitem[Yakowitz et~al., 1978]{yakowitz1978weighted}
Yakowitz, S., Krimmel, J., and Szidarovszky, F. (1978).
\newblock Weighted monte carlo integration.
\newblock {\em SIAM Journal on Numerical Analysis}, 15(6):1289--1300.

\end{thebibliography}
\end{document}